\documentclass[%
 reprint,
 superscriptaddress,
 amsmath,amssymb,prx,
]{revtex4-2}

\usepackage{graphicx}
\usepackage{dcolumn}
\usepackage{bm}
\usepackage{hyperref}
\usepackage{xcolor}
\usepackage{physics}

\def \rucl{$\alpha$-RuCl$_3$}
\def\REE {{\mathrm{Re} }}
\def\IMM {{\mathrm{Im} }}

\newcommand{\nn}{\nonumber \\}
\renewcommand{\parallel}{\mathrel{/\mkern-4mu/}}

\begin{document}

\title{Weak-coupling to strong-coupling quantum criticality crossover in a Kitaev quantum spin liquid $\alpha$-RuCl$_3$}

\author{Jae-Ho Han}
\thanks{These authors contributed equally.}
\affiliation{MPPHC-CPM, Max Planck POSTECH/Korea Research Initiative, Pohang 37673, Republic of Korea}
\affiliation{Department of Physics, Pohang University of Science and Technology, 37673, Republic of Korea}
\affiliation{Asia Pacific Center for Theoretical Physics (APCTP), Pohang 37673, Republic of Korea}
\affiliation{Center for Theoretical Physics of Complex Systems, Institute for Basic Science (IBS), Daejeon, 34126, Republic of Korea}

\author{Seung-Hwan Do}
\thanks{These authors contributed equally.}
\affiliation{MPPHC-CPM, Max Planck POSTECH/Korea Research Initiative, Pohang 37673, Republic of Korea}
\affiliation{Department of Physics, Chung-Ang University, Seoul 06974, Republic of Korea}
\affiliation{Materials Science and Technology Division, Oak Ridge National Laboratory, Oak Ridge, Tennessee 37831, USA}

\author{Kwang-Yong Choi}
\affiliation{Department of Physics, Chung-Ang University, Seoul 06974, Republic of Korea}
\affiliation{Department of Physics, Sungkyunkwan University, Suwon 16419, Republic of Korea}

\author{Sang-Youn Park}
\affiliation{MPPHC-CPM, Max Planck POSTECH/Korea Research Initiative, Pohang 37673, Republic of Korea}

\author{Jae-You Kim}
\affiliation{MPPHC-CPM, Max Planck POSTECH/Korea Research Initiative, Pohang 37673, Republic of Korea}
\affiliation{Division of Advanced Materials Science, Pohang University of Science and Technology, Pohang 37673, Republic of Korea}

\author{Sungdae Ji}
\email{sungdae@kaeri.re.kr}
\affiliation{MPPHC-CPM, Max Planck POSTECH/Korea Research Initiative, Pohang 37673, Republic of Korea}
\affiliation{Neutron Science Center, Korea Atomic Energy Research Institute, Daejeon 34057, Republic of Korea}

\author{Ki-Seok Kim}
\email{tkfkd@postech.ac.kr}
\affiliation{MPPHC-CPM, Max Planck POSTECH/Korea Research Initiative, Pohang 37673, Republic of Korea}
\affiliation{Department of Physics, Pohang University of Science and Technology, 37673, Republic of Korea}

\author{Jae-Hoon Park}
\email{jhp@postech.ac.kr}
\affiliation{MPPHC-CPM, Max Planck POSTECH/Korea Research Initiative, Pohang 37673, Republic of Korea}
\affiliation{Department of Physics, Pohang University of Science and Technology, 37673, Republic of Korea}
\affiliation{Division of Advanced Materials Science, Pohang University of Science and Technology, Pohang 37673, Republic of Korea}

\date{\today}

\begin{abstract}
We report an unprecedented quantum criticality crossover representing two different universal scaling behaviors in a Kitaev quantum magnetic material \rucl{}. \rucl{} presents both a symmetry breaking antiferromagnetic order and a long-range entangled topological order of a quantum spin liquid, and thus could be a candidate system for a new universality class involving deconfined fractionalized excitations of the local Z$_2$ fluxes and itinerant Majorana fermions. Theoretical analyses on the inelastic neutron scattering and specific heat results demonstrate that Wilson-Fisher-Yukawa-type ‘conventional’ weak-coupling quantum criticality in high energy scales crosses over to heavy-fermion-type ‘local’ strong-coupling one in low energy scales. Our findings provide deep insight on how the quantum criticality evolves in fermion-boson coupled topological systems with different types of deconfined fermions.
\end{abstract}

\maketitle

\section{INTRODUCTION}

Searching for a universality class near a quantum critical point (QCP), a precarious point of quantum instability between two competing phases, is a fundamental paradigm with emergence of exotic elementary excitations.
Quantum criticality involving quantum instability \cite{Cole05, Sach08} has been demonstrated in various quantum systems with two competing orders such as magnetic heavy-fermion materials \cite{Schr00}, high-$T_{\mathrm{C}}$ superconductors \cite{Keim15}, one dimensional Ising systems \cite{Cold10}, and Luttinger liquids \cite{Kohn07, Lake05}. Meanwhile, topological orders with emerging gauge fields develop new ground states without symmetry breaking as observed in fractional quantum Hall liquids \cite{Frad13, Stor99} and quantum spin liquids (QSLs) \cite{Bale10, Sava17}, and the low energy physics is described by deconfined fractionalized excitations. In particular, the topological QSL state is exactly derived by fractionalizing the spin excitations into localized Z${}_{2}$ gauge fluxes and itinerant Majorana fermions (MFs) in a two dimensional (2D) Kitaev honeycomb 1/2-spin network with Ising-like nearest-neighbor bond directional Kitaev exchange interactions \cite{Kita06}. Recently, \rucl{} has been found to host both the 2D Kitaev model with fractionalized excitations and a zigzag-type antiferromagnetic (AFM) order below $T_{\mathrm{N}} \approx 6.5\ \mathrm{K}$ \cite{Bane16, John15, Plum14, Sear15}.

The Ru${}^{3+}$ ion in \rucl{} has a $J_{\mathrm{eff}}$ = 1/2 pseudospin \cite{BJKim08} due to strong Ru 4\textit{d} spin-orbit coupling, and the system becomes a spin-orbit coupled Mott insulator \cite{Plum14}. The orbital state forms three orthogonal bonds in the honeycomb lattice to embody the Kitaev model in the edge-shared octahedral environments \cite{Jack09}. Indeed, magnetic susceptibility, specific heat, Raman spectroscopy, and neutron scattering measurements consistently demonstrate the characteristic behaviors of thermally fractionalized MFs above $T_{\mathrm{N}}$, suggesting proximity to the Kitaev QSL phase competing with the AFM phase \cite{Bane17, SHDo17, Sand15, Widm19}. On the other hand, low temperature thermal Hall measurements demonstrated magnetic field driven half-integer quantization representing the quantum nature of fractionalized MFs as the field becomes large enough to suppress the AFM order \cite{Kasa18}.

Meanwhile, the dynamic spin susceptibility, which agrees well to the pure Kitaev behavior in a high energy scale ($T$ or $\omega \gtrsim 5\ \mathrm{meV}$), considerably deviates from the Kitaev behavior in a low energy scale ($T$ or $\omega \lesssim 4\ \mathrm{meV}$) above $T_{\mathrm{N}}$ \cite{SHDo17}. Such proximity behaviors may indicate that the system is nearby a QCP between competing topological QSL and long-range ordered AFM phases and possibly offers a new universality class relevant to the topological order \footnote{Strictly speaking, the phrase ``topological order'' is incorrect because the topological order can be defined only for the gapped systems, while \rucl{} is gapless. X. Wen [Xiao-Gang Wen, Phys. Rev. B \textbf{65}, 165113 (2002)] proposed a term of quantum order based on the concept of projective symmetry group for a gapless phase of matter, but this classification scheme is far from completeness and it seems not widely used in the community. Here, we used the ``topological order'' to point out that the system has {\it fractionalized excitations} which is one of the main features of topological systems in the strict sense.}.

In this study, we report a quantum criticality crossover in \rucl{}, which represents two different universal scaling behaviors in the low and high energy scales. We demonstrate experimentally as well as theoretically that the low-energy spin dynamics obtained from the inelastic neutron scattering (INS) follows heavy-fermion-type strong-coupling physics although the high energy one does weak-coupling physics close to the pure Kitaev. We also ensured that the strong-coupling local quantum criticality at low energies emerges from the weakly-coupled rather conventional Wilson-Fisher-Yukawa-type quantum criticality at high energies. The crossover behavior is also confirmed in the magnetic specific heat $C_{\rm m}$, which displays a curious plateau up to $\sim$ 50 K above $T_{\rm N}$ and then follow a $T$-linear behavior to $\sim$ 100 K, as well explained with the low energy strong-coupling local critical physics and the high energy weak-coupling itinerant Dirac fermion one.

The rest paper consists of as follows. We first propose a physical picture describing a novel quantum criticality crossover emerging from competing Kitaev QSL and zig-zag Ising AFM exchange interactions in Sec. \ref{sec:phys_pic}. Sec. \ref{sec:INS}  presents INS results representing the dynamic spin susceptibility. In Sec. \ref{sec:EFT}, we discuss the effective model Hamiltonian and field theory description. Then we derive the theoretical universal scalings for the susceptibility compared with the experiments in Sec. \ref{sec:high_en}. Weak Coupling High-Energy Region and in Sec. \ref{sec:low_en}. Strong Coupling Low-Energy Region. In Sec. \ref{sec:mag_sph}, we discuss the weak to strong coupling crossover in the magnetic specific heat. Finally, we summarize and discuss the present study in Sec. \ref{sec:summary}. Additional information and details are provided in Appendixes and Supplemental Materials. 

\section{Physical picture}
\label{sec:phys_pic}

\begin{figure}[t]
\includegraphics[width=8.5cm]{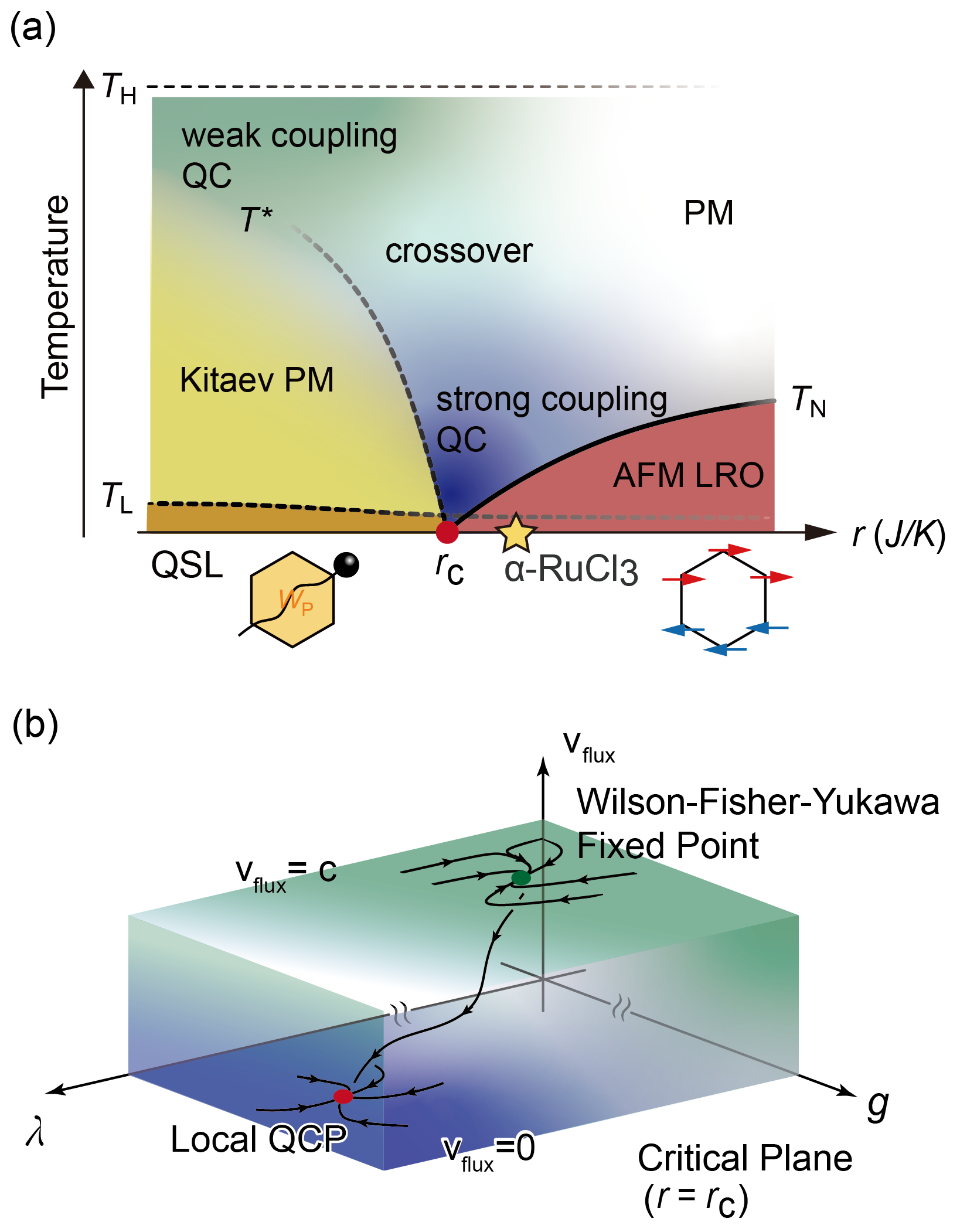}
\caption{
(a) Schematic phase diagram as a function $r$ ($J/K$), a ratio of the non-Kitaev antiferromagnetic (AFM) exchange $J$ to the Kitaev exchange $K$. The quantum critical region is divided into a weak-coupling one in high temperatures (green-shaded) and a strong-coupling one in low temperatures (blue-shaded) governed by different critical scaling physics, resulting in a novel crossover behavior of the quantum criticality (QC). (b) Renormalization group (RG) flows on the $r=r_{\mathrm{C}}$ critical surface described with two coupling constants $\lambda$, $g$, and velocity $v_{\mathrm{flux}}$ of the Z$_2$ flux excitation. The RG flows at high temperatures ($v_{\mathrm{flux}} = c$) show actual results of the Wilsonian RG analyses in the one-loop level while those are schematically presented at low temperatures ($v_{\mathrm{flux}} = 0$).}
\label{fig:phase_diagram}
\end{figure}

We first discuss a physical picture with respect to a novel universality class and quantum criticality for critical spin fluctuations in a model system with two competing interactions of 2D Kitaev and Ising-type AFM exchange interactions as in \rucl{}.The AFM spin fluctuations couple to two different types of deconfined fractional excitations of localized Z$_2$ fluxes and itinerant MFs dominantly in low and high energy scales, respectively.

At high temperatures far above $T_{\mathrm{N}}$, the Z${}_{2}$ fluxes and MFs dissolve to form an incoherent critical soup. The original quasiparticle weights become vanishing, and emerging excitations losing the quasiparticle nature specify characteristics of weak-coupling conventional quantum criticality. This phenomenon is analogous to physics of metallic quantum criticality responsible for the marginal Fermi liquid phenomenology in the strange metallic phase of high $T_{\mathrm{C}}$ cuprates \cite{Keim15}, where the Landau quasiparticle excitations dissociate into incoherent particle-hole excitations due to their correlations. Remarkably, this conventional quantum criticality evolves into strong-coupling local quantum criticality at low temperatures above $T_{\mathrm{N}}$, where critically fluctuating local moments appear to carry the entropy, analogously to the local quantum criticality of heavy-fermion systems \cite{Schr00}. This crossover behavior in quantum critical physics of \rucl{} originates from coupling between the Ising-type AFM fluctuations and the two types of deconfined fractionalized fermions governing different energy scales.

Figure \ref{fig:phase_diagram}(a) shows a schematic phase diagram representing the Kitaev QSL phase with the well-defined fractionalized excitations and the zigzag AFM phase, similarly to the diagram of the Landau Fermi-liquid phase with well-defined quasiparticle excitations and an associated symmetry broken phase, respectively. Here the external tuning parameter such as the magnetic field or pressure varies the ratio of the AFM exchange $J$ to the QSL Kitaev exchange $K$. In a $J=0$ limit, the spin-spin correlation is governed by $K$ leading to the Kitaev QSL ground state. The spin excitations are fractionalized into itinerant MFs and gapped Z${}_{2}$ fluxes, which become consecutively defined upon cooling through $T_{\mathrm{H}}$ and $T_{\mathrm{L}}$, respectively \cite{Nasu15, Yosh16}. 

In the Kitaev paramagnetic (PM) phase ($T_{\mathrm{L}} < T < T_{\mathrm{H}}$), the MFs exhibit metallic behaviors under thermally fluctuating Z${}_{2}$ fluxes. Below $T_{\mathrm{L}}$, the Z${}_{2}$ fluxes are frozen, and only low energy MFs are itinerant. Even for a finite $J$, the QSL ground state persists if $r = J/K$ is sufficiently small. As $r$ increases, the AFM coherence length ($\xi$) increases and the coherence temperature $T^* \sim {\xi }^{-1}$ decreases. In a large $r$ limit, the AFM long-range order (LRO) is stabilized below $T_{\mathrm{N}}$ due to the dominant $J$. The quantum phase transition between QSL and LRO phases is expected to occur at a moderate $r=r_{\rm C}$ (QCP), and the physical behaviors become quantum critical in the region (blue-shaded) around QCP. Considering that the magnetic field induced AFM to QSL transition occurs at $H_{\rm C} \sim 6$ T (see Fig. \ref{fig:bulk_property}) \cite{Kasa18} in \rucl{}, its $r$-value is expected to be near the critical point $r_C$ as presented in the figure.

The most fascinating physics is an apparent crossover behavior across the high (green-shaded) to low temperature (blue-shaded) region, suggesting two types of $\omega /T$ quantum critical scaling physics in the dynamic spin susceptibility. This crossover behavior results from the fact that the spin excitations in high and low energy scales are governed by different fixed points (FPs) and there appears a renormalization group (RG) flow between these two FPs as a function of the energy scale parameter ($T$ or $\omega$). As a result, the fractionalized excitations form a novel QSL state in this quantum critical region. Here the phase boundaries are referred to experimental observations \cite{SHDo17, Sand15, Widm19, Kasa18} and theories \cite{Nasu15, Yosh16} although the diagram is rather schematic. The crossover region in Fig. \ref{fig:phase_diagram}(a) is referred to the neutron scattering (Fig. \ref{fig:neutron_data}(b)) and magnetic specific heat results (Fig. \ref{fig:scaling_sph}(a)).

Based on the physical picture described in Fig. \ref{fig:phase_diagram}(a), we explore the nature of this quantum criticality, where the critical AFM spin fluctuations couple to both Z${}_{2}$ fluxes and MFs. Figure \ref{fig:phase_diagram}(b) displays the RG flow diagram around the critical surface at $r = r_{\mathrm{C}}$. The RG flows in the high energy scale show actual calcluation results of the Wilsonian RG analyses in the one-loop level on the critical surface. Meanwhile those are schematically presented in the low energy scale, where the physics involves strong-coupling between the deconfined excitations and AFM fluctuations to be beyond the perturbative framework. However, it is worth to note that the existence of the strong-coupling FP is verified by the dynamical mean-field theory (DMFT) calculation as discussed in Sec. \ref{subsec:DMFT}.

In absence of fractionalized excitations, the Ising spin quantum criticality is represented with the Wilson-Fisher type FP \cite{Zinn02} characterized by an effective critical interaction $\lambda$ between the Ising spin fluctuations. Meanwhile, as this spin sector interaction is turned off ($\lambda = 0$), the fractionalized fermionic spin excitations can be described in terms of an effective Yukawa-type theory characterized by an effective critical Yukawa coupling $g$ between the fractionalized excitations and the spin fluctuations. This theory naturally hosts a weak-coupling FP \cite{Zinn02}. As $\lambda \neq 0$, an interacting FP, referred to as a Wilson-Fisher-Yukawa FP, emerges on the critical surface as presented in Fig. \ref{fig:phase_diagram}(b) (the high temperature green-shaded region in Fig. \ref{fig:phase_diagram}(a)). 
 
The proximity effect observed in \rucl{} at high temperatures is expected to be governed by this interacting FP. The dynamic spin susceptibility should exhibit a universal behavior and further follow a universal scaling law characterized by an anomalous critical exponent derived from this FP. It is remarkable to note that this Wilson-Fisher-Yukawa FP becomes destabilized at low temperatures ($T \lesssim 4\ \mathrm{meV}$) above $T_{\mathrm{N}}$ to flow into another novel FP, where the propagation velocity of the Z${}_{2}$ flux excitation is strongly renormalized (localized Z${}_{2}$ flux excitation). Then the nature of spin dynamics becomes locally critical (a strong-coupling FP) as described for the local quantum criticality of heavy-fermion physics. Appearance of the localized flux excitations in the Kitaev QSL state is likely responsible for the heavy-fermion type local quantum criticality in the low energy scale.

\section{Inelastic neutron scattering}
\label{sec:INS}

\begin{figure}[t]
\includegraphics[width=8.5cm]{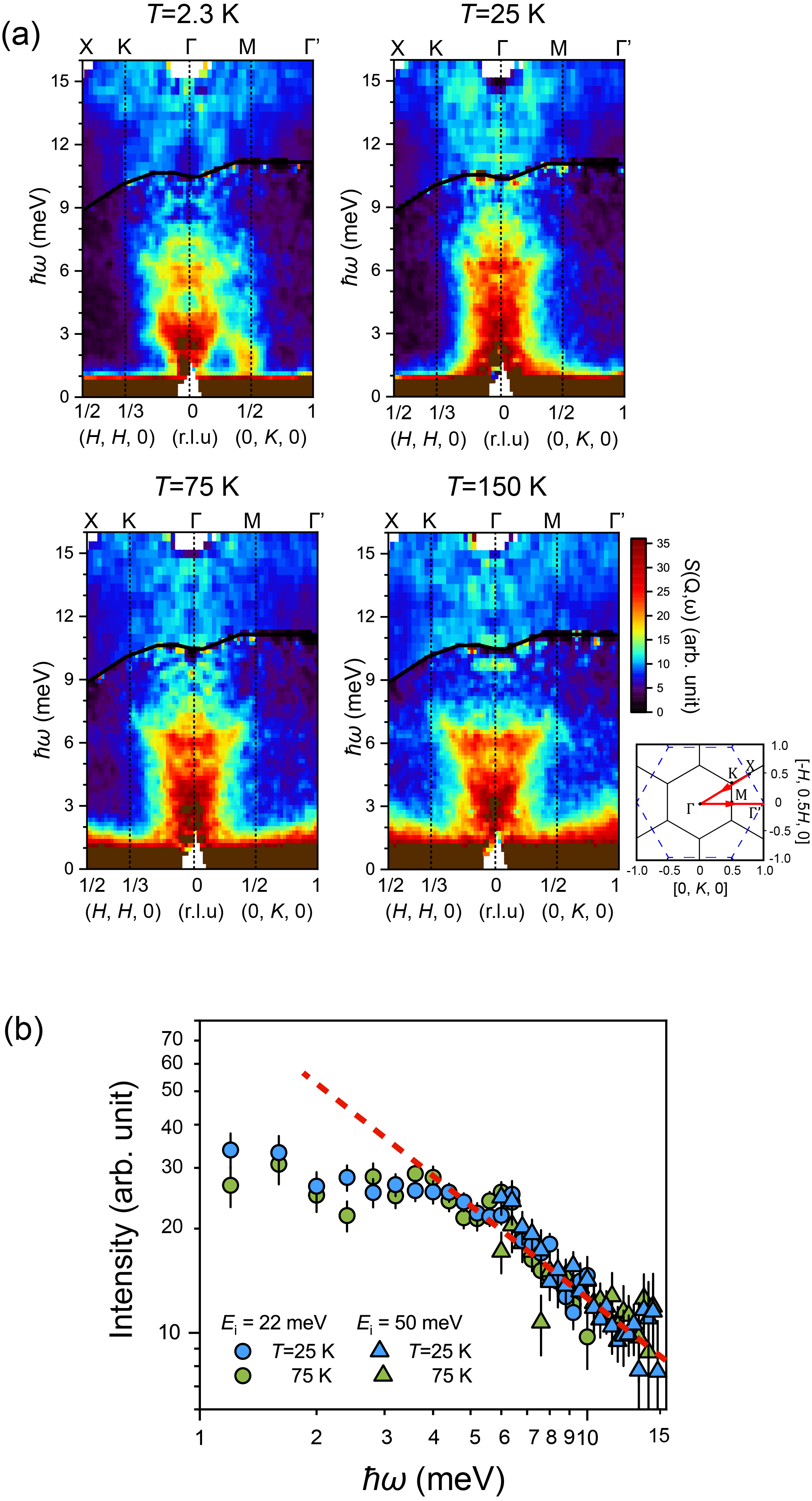}
\caption{
(a) Inelastic neutron scattering $S(\bm Q,\omega)$ maps along high symmetry directions as indicated in the $HK$-reciprocal space of the rhombohedral ($R \bar 3$) structure shown in (c). Each map presents combined two data sets with incident neutron energies of $E_\text{i} = 50 \ {\rm meV}$ (upper) and $22 \ {\rm meV}$ (lower).
(b) $S(\bm Q,\omega)$ spectra at $\bm Q = \Gamma \ (0,0,0)$ measured at $T$ = 25$\ \mathrm{K}$ and $75\ \mathrm{K}$ in a log-log scale. Red dashed line is a linear guideline for a power-law behavior of the excitations above $\sim$ 4 meV.}
\label{fig:neutron_data}
\end{figure}

The dynamic spin susceptibility can be extracted from INS measurements exploring the magnetic excitations. The scattering cross-section is proportional to the dynamic structure factor $S(\bm Q,\omega)$ of the spin correlation function, which displays the magnetic excitations in the transferred momentum ($\bm Q$) and energy ($\omega$) space. Figure \ref{fig:neutron_data}(a) shows representative $S(\bm Q,\omega)$ maps of \rucl{} in the $HK$-reciprocal space measured at below ($T = 2.3\ \mathrm{K}$) and above $T_{\mathrm{N}} = 6.5\ \mathrm{K}$ ($T = 25\ \mathrm{K}$, $75\ \mathrm{K}$, and $150\ \mathrm{K}$). The $S(\bm Q,\omega)$ map at $T=2.3\ \mathrm{K}$ exhibits a strongly dispersive feature of spin wave excitations originated from the zigzag AFM order below $4\ \mathrm{meV}$. Besides the spin wave feature, it also displays broad continuum excitations extending from near $0\ \mathrm{meV}$ to even above $15\ \mathrm{meV}$ through the whole Brillouin zone, which correspond to the fractionalized MFs excited from the Kitaev QSL state \cite{SHDo17, Yosh16}. These continuum excitations emerge upon cooling across the fractionalization temperature ($T_{\mathrm{H}} \sim 100\ \mathrm{K}$) and persist even below $T_{\mathrm{N}}$, reflecting proximity to the QSL \cite{SHDo17}.

Above $T_{\mathrm{N}}$ ($25\ \mathrm{K}$ and $75\ \mathrm{K}$), one can recognize distinguishable energy dependent behaviors of the continuum excitation weight around $\bm Q$=$\Gamma$(0,0,0) across $\sim 4\ \mathrm{meV}$; a strong spectral weight is nearly maintained up to $\sim 4\ \mathrm{meV}$ and then become significantly reduced with increase of the energy. Such energy dependent behaviors can be clearly observable in the $S(\bm Q$ = $\Gamma$,$\omega)$ spectra in the log-log scale as shown in Fig. \ref{fig:neutron_data}(b). Above $\sim 4\ \mathrm{meV}$, the spectral intensity merges on a linear line representing a single scaling of criticality while it apparently deviates from the line below $\sim 4\ \mathrm{meV}$, speculating emergence of another scaling. This result is consistent with the weak-coupling to strong-coupling quantum criticality crossover as described above (also see Fig. \ref{fig:phase_diagram}) and suggests that the nature of continuum excitations can be characterized by two different universal scaling laws, i.e. one for the weak-coupling and the other for the strong-coupling quantum criticality in the high and low energy scales, respectively.Interestingly, a puzzling plateau feature above $T_{\rm N}$ observed in the specific heat \cite{SHDo17} persists up to $50 \ {\rm K}$ corresponding to the deviation energy scale ($\sim 4 \ {\rm meV}$). We speculate that the delocalized $Z_2$ flux excitations at high temperatures become localized below this temperature although it is well above the freezing temperature $T_{\rm L}$ in the pure Kitaev \cite{Nasu15, Yosh16}, causing a heavy-fermion type local quantum criticality as discussed below (Sec. \ref{sec:low_en}).

\section{Effective field theory}
\label{sec:EFT}

In this section, we introduce an effective field theory modeling for the spin excitations of \rucl{} with the AFM spin fluctuations coupling to the $Z_2$ fluxes and itinerant MFs in order to derive the universal scaling applicable to the dynamic spin susceptibility extracted from the INS results. The true magnetic lattice Hamiltonian of \rucl{} is quite complicated and still under debate. Thus we build up a simplified Kitaev-QSL interacting with the zigzag AFM order in the \rucl{} lattice.

\begin{figure}[t]
\includegraphics[width=8.5cm]{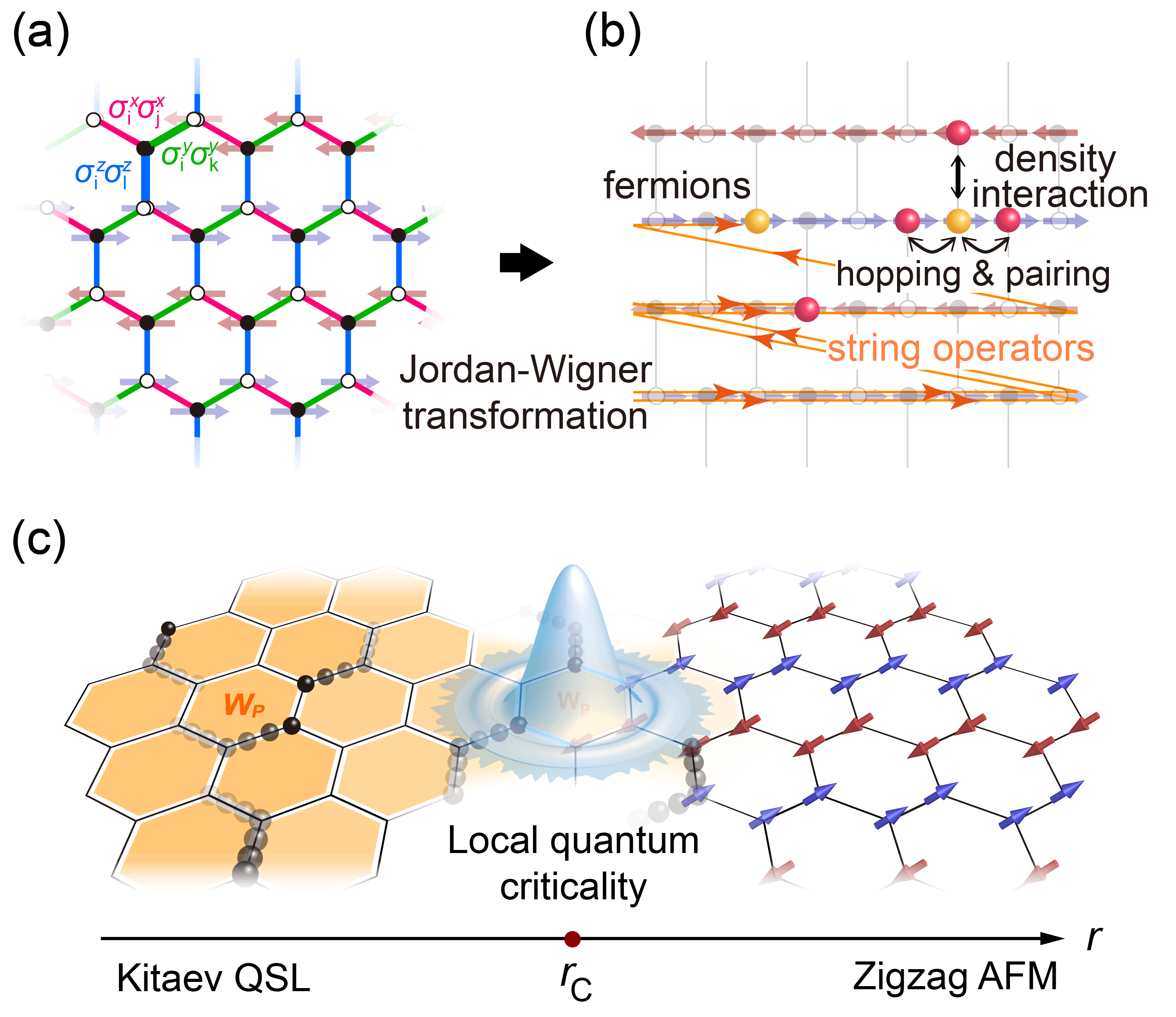}
\caption{Schematic descriptions of the theoretical model and quantum critical scaling behaviors of dynamic spin susceptibility.
(a) Kitaev honeycomb lattice represented by two triangular sublattices differentiated by black and white dots and zigzag antiferromagnetic ordering.
(b) Deformed brick-wall type lattice by stretching the honeycomb lattice along the zigzag chain. The red and yellow balls are $f_{r,b}$ and $f_{r,w}$ fermions, respectively.
(e) Conceptual diagram for emergent local quantum criticality in the low energy scale. }
\label{fig:EFT}
\end{figure}

We first adopt a coarse-grained lattice model (Fig. \ref{fig:EFT}(a)) to construct the effective Hamiltonian consisting of the Kitaev QSL Hamiltonian $H_{\mathrm{K}}$, the zigzag AFM Ising spin Hamiltonian $H_{\mathrm{AF}}$, and an effective Zeeman-type interaction $V_{\mathrm{K-AF}}$ accounting for a coupling between the Kitaev fermions and the zigzag AFM ordered spins. The total Hamiltonian can be described as follows,
\begin{align}
H &= H_{\mathrm{K}} + H_{\mathrm{AF}} + V_{\mathrm{K-AF}}, \nn
H_{\mathrm{K}} &= - \sum_{\left< ij \right>} K_{\gamma_{ij}} \sigma_i^{\gamma_{ij}} \sigma_j^{\gamma_{ij}}, \ \ \
V_{\mathrm{K-AF}} = - g \sum_i \phi_i \sigma_i^z,
\end{align}
where $\sigma_i^\alpha$, $\alpha = x, y, z$ are Pauli matrices at honeycomb lattice site $i$, $\left< ij \right>$ represent nearest-neighboring sites, $\gamma_{ij} = x, y, z$ depending on the bond as shown in Fig. \ref{fig:EFT}(a), and $\phi_i$ is the zigzag AFM order parameter fluctuations. Here, we do not specify $H_{\mathrm{AF}}$ explicitly due to the lack of lattice magnetic Hamiltonian in consensus. The detailed form of $H_{\rm AF}$ is not important in this study only for the long-range physics. We will implement dynamics of the zigzag Ising AFM order in a continuum expression form when we construct the effective action below. The interaction $V_{\rm K-AF}$ between $\sigma$ and $\phi$ is supposed to be in a local Zeeman type.

The zigzag AFM order consists of alternating spin chains with different spin states, and the original honeycomb lattice is identically represented by a brick-wall lattice. Under the Jordan-Wigner transformation for the spin variables described in Fig. 3(b), one can map the Kitaev spin model to a fermionic model: Transforming the spin operators $\sigma_i^\alpha$ as
\begin{align}
\sigma_i^z = K_i, \ \sigma_i^+ = f_i^\dagger \prod_{j<i} K_j , \ \sigma_i^- = \bigg(\prod_{j<i} K_j \bigg) f_i,
\end{align}
where $K_i = 2f_i^\dagger f_i -1$, and $f_i$ ($f_i^\dagger$) are fermion annihilation (creation) operators. Then $H_{\mathrm{K}}$ becomes
\begin{align}
H_{\mathrm K} =& - \sum_{r, a=x,y}K_a (\psi_{r+a}^\dagger - \psi_{r+a}) (\psi_r^\dagger + \psi_r) \nn
& -K_z \sum_r (2\chi_r^\dagger \chi_r - 1) (2\psi_r^\dagger \psi_r- 1),
\end{align}
where the lattice site index $i$ are separated into Bravais lattice index $r$ and the basis $b$ or $w$ for black or white site respectively, shown in Fig. \ref{fig:EFT}. We also combine the fermions attached to two ends of the $z$-bond as $\chi_r = \REE f_{r,w} - i\IMM f_{r,b}$, and $\psi_r = \IMM f_{r,w} + i\REE f_{r,b}$. Now $\psi_r$'s resemble the itinerant fermions in the one-dimensional $p$-wave superconductor model and $\chi_r$'s representing the $Z_2$ fluxes describe localized fermions interacting with $\psi_r$ fermions \cite{Zinn02, DHLee07}.

Now we consider $V_{\rm K-AF}$, the coupling between the Ising spin fluctuations and the Jordan-Wigner fermions in the critical region. In the coarse-grained representation, this coupling can be expressed as a Yukawa-type effective interaction, and the model Hamiltonian becomes an effective continuum field theory described in terms of two types of fermion excitations coupled with Ising spin fluctuations. The Yukawa-type effective interaction is given by
\begin{align}
V_{\mathrm{K-AF}} =& - ig \sum_r \phi_i (\psi_r^\dagger \chi_r^\dagger + \psi_r \chi_r) .
\end{align}

To obtain the low-energy effective field theory, we expand both fermion fields near two gapless points in the momentum space and construct spinor fields in a standard manner. Then, the resulting effective action is in a Dirac form, given by
\begin{align}
S =& S_{\mathrm K} + S_{\mathrm{AF}} + S_{\mathrm{K-AF}}, \nn
S_{\mathrm K} =& \int\!\frac{d^3k}{(2\pi)^3} \ \Big( \bar\Psi_k i\gamma \cdot k \Psi_k + \bar X_k (i \gamma_{0} k_{0} + i\gamma \cdot M) X_k \Big), \nn
S_{\mathrm{K-AF}} =& -ig \int\!\frac{d^3k d^3q}{(2\pi)^6} \ \phi_q \big( \bar X_{k+q} \Psi_k - \bar\Psi_k X_{k-q} \big), \nn
S_{\mathrm{AF}} =& \frac{1}{2} \int\!\frac{d^3k}{(2\pi)^3} \ \phi_{-k} (k^2 + r) \phi_k \nn
&+ \frac{\lambda}{4!} \int\!\frac{d^3k d^3p d^3q}{(2\pi)^9} \ \phi_{-k} \phi_{-p-q} \phi_p \phi_{k+q}.
\label{eq:eff_action}
\end{align}
Here, $\Psi_k$ and $X_k$ are Dirac spinors formed by itinerant and localized fermions, respectively, and $M$ is the $Z_2$ flux gap. Interestingly, the Z${}_{2}$ flux gap appears as a momentum shift of nodal points, and this shift plays a central role in the quantum critical scaling of dynamic spin susceptibility at the $\mathrm{\Gamma }$ point. There appears an effective $\phi^4$-type field theory of self-interactions \cite{Zinn02}, which describes critical dynamics of the Ising spin fluctuations, in the coarse graining procedure. The details are in Supplemental Material A1.

In the following sections, we derive the scaling behavior of the dynamic spin susceptibility (Sec. \ref{sec:high_en} and \ref{sec:low_en}) and the specific heat (Sec. \ref{sec:mag_sph}) based on Eq. (\ref{eq:eff_action}) in the weak-coupling limit for the high-energy region and also in the strong-coupling limit for the low-energy region (Sec. \ref{sec:low_en}).

\section{Weak coupling high-energy region}
\label{sec:high_en}

In the high-energy region, the coupling becomes weak so that we derive the universal scaling function using the perturbative analysis. The zigzag AFM order parameter also strongly fluctuates and thus the ``localized'' $X_k$ fermions becomes delocalized due to the interaction $S_{\rm K-AF}$. As the result, the two fermionic excitations can be described in the Dirac theory with different Dirac velocities. We first perform the Wilsonian RG analyses of this two itinerant Dirac field model up to the one-loop level and reveal the Wilson-Fisher-Yukawa FP. Then we evaluate the dynamic spin susceptibility at the FP for comparison with the INS results of \rucl{}.

\subsection{Wilson-Fisher-Yukawa-type fixed point}

\begin{figure}[t]
\includegraphics[width=8.5cm]{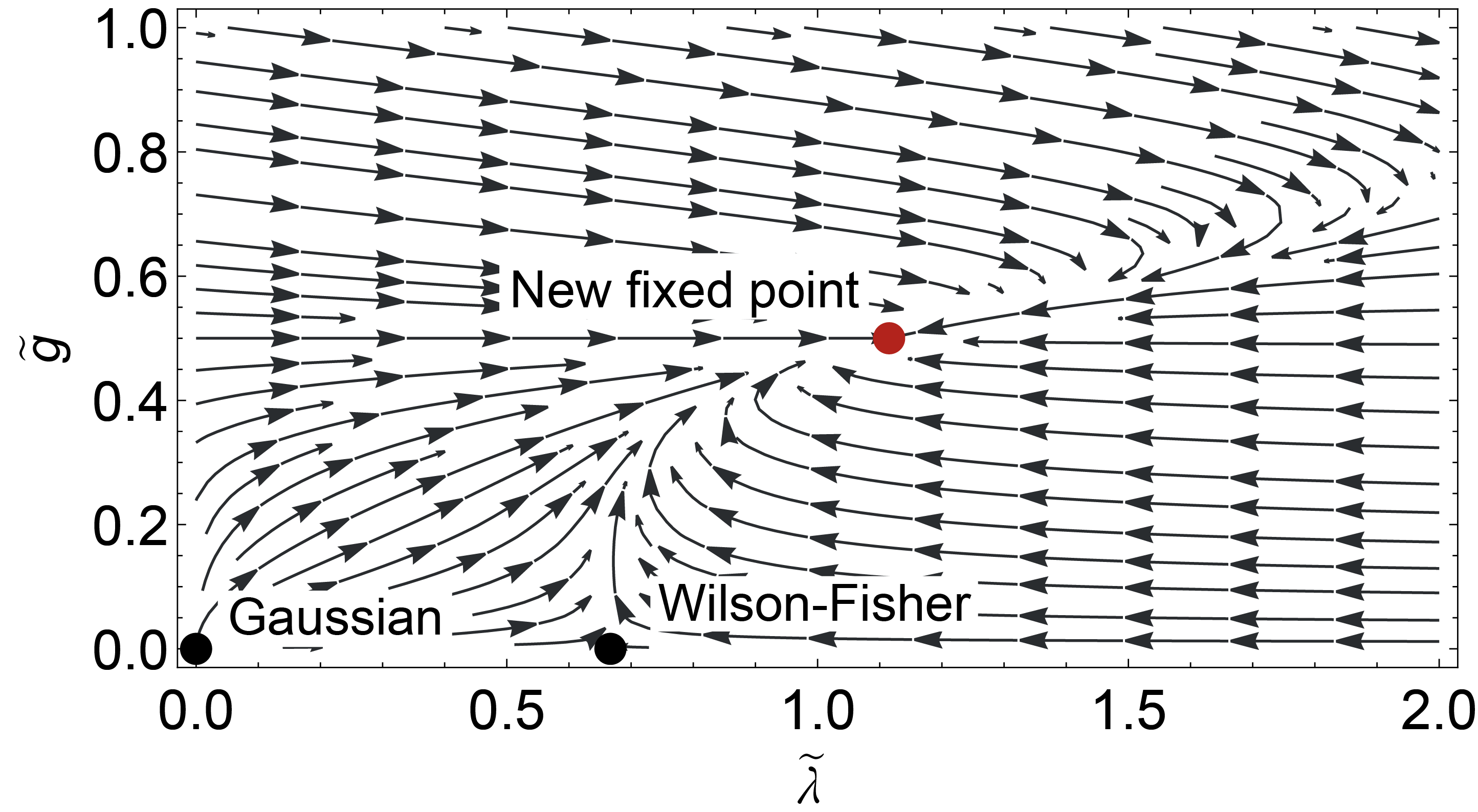}
\caption{The renormalization group flow diagram of coupling constants. Solving the coupled $\beta$-functions, we find RG flows in a $\tilde g$-$\tilde\lambda$ coupling space with $\tilde g^2 = g^2/8\pi^2$ and $\tilde\lambda = \lambda/8\pi^2$, and obtain a novel interacting fixed point (red dot).}
\label{fig:RG_flow}
\end{figure}

Reflecting delocalization of the $X_k$ fermions, we replace $\gamma_0 k_0$ with $\gamma \cdot k$ to modify the $S_{\rm K}$ term in Eq. (\ref{eq:eff_action}) as follows;
\begin{align}
S_{\mathrm K} \rightarrow& \int\!\frac{d^3k}{(2\pi)^3} \ \Big[ \bar\Psi_k i\gamma \cdot k \Psi_k + \bar X_k i\gamma \cdot (k+M) X_k \Big].
\end{align}
With the modified effective action, we follow the standard Wilsonian RG precedure. We first, divide the fields into high- and low-energy components, and integrating out the high-energy parts. After rescaling the momentum coordinates and all low-energy fields, we obtain an effective action for the low-energy fields with renormalized couplings. Such renormalized couplings are described by the following beta functions:
\begin{align}
\beta_g &\equiv \frac{dg}{d\log b} = \frac{g}{2} - \frac{g^3}{4\pi^2} + \mathcal O(M), \nn
\beta_\lambda &\equiv \frac{d\lambda}{d\log b} = \lambda - \frac{3\lambda^2}{16\pi^2} + \frac{3g^4}{2\pi^2} + \mathcal O(M),
\end{align}
where $b$ is a scaling parameter (see Supplemental Material A2). The RG flow and the FPs are shown in Fig. \ref{fig:RG_flow}. As can be seen in the figure, when $g=0$, the fermions are decoupled to the zigzag AFM fluctuations and the system flows to a well-known stable Wilson-Fisher FP. Turning on the coupling ($g\neq 0$) between fermions and zigzag AFM fluctuations, it flows to a new FP, Wilson-Fisher-Yukawa FP. Essential information on the interacting FP is critical exponents to describe anomalous scaling dimensions of the delocalized and itinerant fermion excitations, and the exponents originate from their correlations with the critical Ising spin fluctuations.

To compare with INS data at $\Gamma$ point, we calculate uniform spin susceptibility. At the FP of the beta functions, one should differentiate the effective renormalized partition function with respect to the external magnetic field. We include the effect of the renormalization after the integration of the internal momentum. Then one can obtain the following universal scaling function at the $\mathrm{\Gamma}$ point in the quantum critical regime of $\omega \gtrsim 1$
\begin{equation}
T^\alpha\ {\mathrm{Im}\chi (\omega ,T)}_{Q=\mathrm{\Gamma }} = \chi_0 M^2 {\left(\frac{T}{\omega }\right)}^{1.25} \mathrm{tanh} \frac{\omega }{4T}.
\label{eq:scaling_highE}
\end{equation}
with $\alpha = 1$. Here ${\mathrm{tanh} \frac{\omega }{4T}}$ reflects `particle'-`hole' excitations of both fractionalized fermions and ${\chi }_0$ is a cut-off dependent non-universal constant. $M$ is the momentum-space distance between the Dirac points of the Majorana fermion and the Z${}_{2}$ flux and reduces to the Z${}_{2}$ flux gap at zero temperature. It is remarkable to observe that the spin susceptibility at the $\mathrm{\Gamma }$-point is proportional to $M^2$. Although the spectral intensity of the two-particle correlation function should vanish at the $\mathrm{\Gamma }$-point with $M = 0$, appearance of the spectral intensity at the $\mathrm{\Gamma }$-point indicates that a shift of the nodal point effectively retains the Z${}_{2}$ flux gap as like the inter-band transition gap.

\subsection{High-energy scaling}

\begin{figure}[b]
\includegraphics[width=8.5cm]{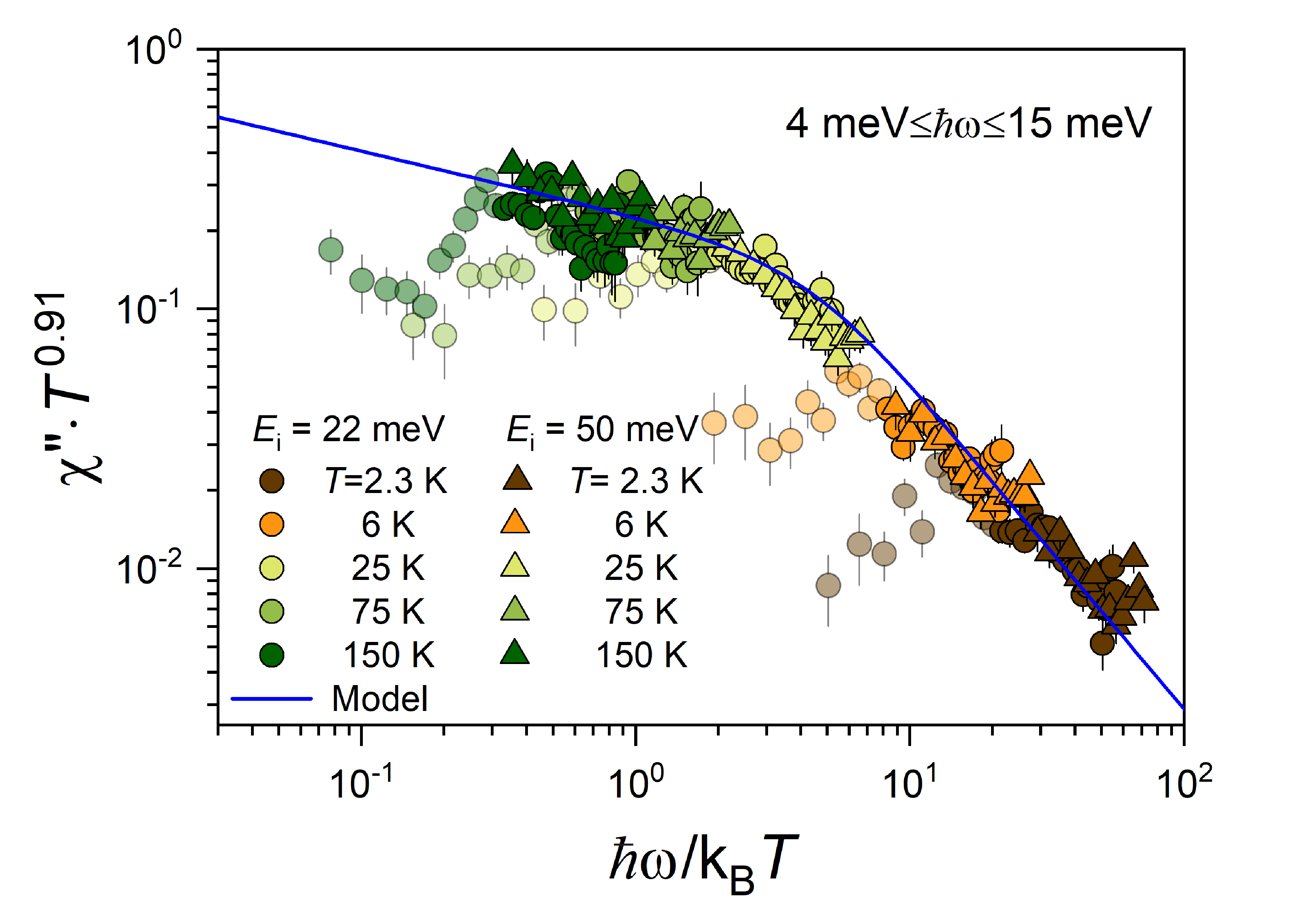}
\caption{Scaling plot in $\hbar\omega / k_BT$ for the dynamic spin susceptibility at high energies, $\hslash \omega \ge 4\ \mathrm{meV}$. Blurred circles present the data in the low energy range $\hslash \omega$ = [1,4] meV, which are out of the universal scaling behavior.}
\label{fig:spectra_high}
\end{figure}


To investigate the critical behavior of \rucl{}, the neutron scattering results at $\bm Q = \mathrm{\Gamma }$ with diverse temperatures are compared to the scaling function model. Here the imaginary part of the dynamic spin susceptibility is extracted through the fluctuation-dissipation theorem \cite{Squi78} $\mathrm{Im}\chi \left(\bm Q,\omega \right)\equiv \chi'' \left(\bm Q,\omega \right)=S(\bm Q,\omega )\left(1-e^{-\hslash \omega /k_BT}\right)$ from $S(\bm Q,\omega )$ measured at $T$ = 2.3 K, 6 K, 25 K, 75 K, 150 K in a full energy range of $1\ \mathrm{meV}<\hslash \omega < 15\ \mathrm{meV}$ available in the present experimental conditions (see Appendix A). Figure \ref{fig:spectra_high} displays $\chi'' (\mathrm{\Gamma }, \omega) T^\alpha$ versus $\hslash\omega /k_BT$ in the log-log plot. The scaling function data extracted from the zero-field INS $S(\Gamma, \omega)$ in the ranges $\hslash \omega$ = [1, 9] meV with incident neutron energy $E_\text{i}$ = 22 meV and $\hslash \omega$ = [6, 15] meV with $E_\text{i}$ = 50 meV at $T$ = 2.3 K, 6 K, 25 K, 75 K, 100 K. The data are compared with the theoretical universal scaling function (blue solid line) for the weak-coupling quantum criticality, Eq.~(\ref{eq:scaling_highE}). The fitting value of the exponent $\alpha$ is $0.91$, which is close to the theoretical value $1$. The $\chi'' (\mathrm{\Gamma }, \omega) T^\alpha$ value itself strongly varies with energy and temperature while those values collapse onto a single line over two decades for $\hslash\omega \gtrsim 5\ \mathrm{meV}$. Such behaviors certainly reflect the universal scaling involving the weak-coupling quantum criticality applicable to the high energy scale. This merging line corresponds to the universal law for $\chi'' (\Gamma, \omega) T^\alpha$ derived from the theoretical model calculations as described above. Meanwhile, one can recognize that the dynamic spin susceptibility does not follow the university scaling at low energies, and the deviation becomes considerable for $\hslash\omega \lesssim 4\ \mathrm{meV}$ commonly at different temperatures as can be seen in Fig. \ref{fig:spectra_high} (blurred circle plots). This common deviation at the low energies indicates that the weak-coupling quantum criticality is valid only at the high energies (also see Fig. \ref{fig:neutron_data_appnd} in Appendix C).

\section{Strong coupling low-energy region}
\label{sec:low_en}

In the low-energy region, the perturbative analysis is not applicable since the strong-coupling physics emerges. Thus we first examine the INS results for a possibility of a universal scaling at the low energies and construct an empirical formula for the dynamic susceptibility in the effective model Hamiltonian with the local quantum criticality. Then we check the validity of the criticality using self-consistent analyses based on the DMFT description.

\subsection{Low-energy scaling: local quantum criticality}

\begin{figure}[t]
\includegraphics[width=8.5cm]{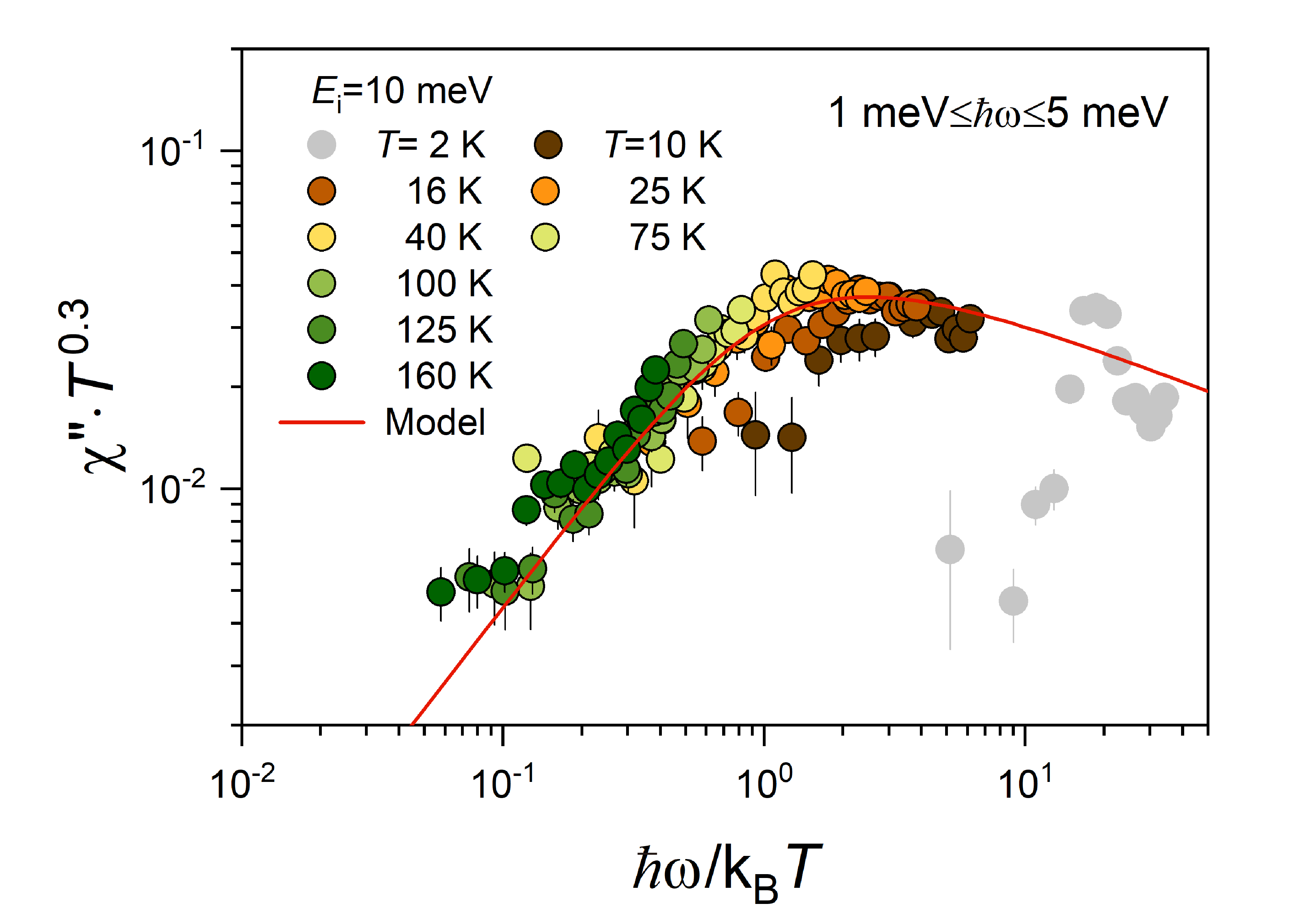}
\caption{Scaling plot for the dynamic spin susceptibility in $\hbar\omega / k_BT$ at low energies, $E \le 5\ \mathrm{meV}$. The grey circles presenting the 2 K data seriously deviate from the universal scaling due to magnon excitations in the AFM state below $T_\mathrm{N}\approx$ 6.5 K.}
\label{fig:spectra_low}
\end{figure}

To examine the low energy universal scaling of the dynamic spin susceptibility, we measure $S(\bm Q,\omega)$ at more diverse temperatures of $T$ = 2 K, 10 K, 16 K, 25 K, 40 K, 75 K, 100 K, 125 K, 160 K in a low energy range of $1\ \mathrm{meV} \le \hslash\omega \le 5\ \mathrm{meV}$ and extract $\chi'' (\mathrm{\Gamma }, \omega) T^{0.3}$ values scaled in $\hslash\omega / k_BT$, as shown in Fig. \ref{fig:spectra_low}. Remarkably, the $\chi'' (\mathrm{\Gamma }, \omega) T^{0.3}$ values also merge to a single line of another universal scaling distinguished from the high energy one. The universal scaling behavior drastically changes in the low energy scale below $\sim 5\ \mathrm{meV}$. The slop changes its sign across $\hslash\omega / k_{B}T \sim 2$. In addition, the low energy spectral weight is rather uniformly distributed in the momentum space around $\bm Q = \Gamma$, differently from the high energy spectral weight (referred to Fig. \ref{fig:spectra_high}). These two aspects recall an effective Bose-Fermi Kondo-type model adopted to the heavy fermion local quantum criticality in a system with magnetic impurity states, collective bosonic modes, and dispersive fermions \cite{QSi01}. Those are analogous to the localized Z${}_{2}$ fluxes, Ising AFM fluctuations, and itinerant Majorana fermions appearing in \rucl{}, respectively. This local quantum criticality is schematically pictured in Fig. \ref{fig:EFT}(c). Here the AFM fluctuation interacts with the Z${}_{2}$ fluxes and MFs to become locally critical.

The previous study for the heavy-fermion local quantum criticality \cite{Schr00} suggests a scaling expression for the dynamic spin susceptibility
\begin{align}
\chi \left(\omega, T, H\right) = \frac{A}{{\left(aT-i\omega \right)}^{\alpha }+a^{\alpha }T^{*\alpha }},
\label{eq:chi_LQCP}
\end{align}
which represents the susceptibility of a local moment coupled to a critical continuum. Here, $a$, $A$, and $\alpha$ are parameters, $T^*$ is the characteristic temperature, shown in Fig. \ref{fig:phase_diagram}(a). A key feature of this local spin susceptibility is the branch-cut singularity with a critical exponent $\alpha$ and existence of the huge dissipation proportional to the transfer energy. We derive the universal scaling function at $\Gamma$ for the strong-coupling local quantum criticality in a limit of the inverse of quantum coherence time $T^*\to 0$ (QCP) as follows (see Supplemental Material B1 for details);
\begin{eqnarray}
&&T^{\alpha }\ {\mathrm{Im} \chi (\omega ,T )}_{\bm Q=\mathrm{\Gamma }} \nn &&\hspace{30pt} = \frac{A}{{\left(a^2T^2+{\omega }^2\right)}^{\alpha /2}} \sin \left\{\alpha \ \tan^{-1}\left(\frac{\omega }{aT}\right)\right\}.
\label{eq:scaling_lowE}
\end{eqnarray}
As can be seen in Fig. \ref{fig:spectra_low}, this theoretical scaling function well explains the low energy universal scaling behavior obtained from the INS results ranging over about two order of magnitude in $\hslash\omega /k_BT$ above $T_{\mathrm{N}}$ with $\alpha \approx 0.30$, the overall constant $A \approx 0.144$, and the order 1 constant $a \approx 0.95$.

\subsection{Dynamical mean-field theory analysis}
\label{subsec:DMFT}

To confirm existence of the local quantum criticality at low temperatures in the vicinity of the genuine QCP, we perform a DMFT analysis for the localized Z${}_{2}$ flux excitations, itinerant MFs, and locally critical Ising AFM spin fluctuations within a non-crossing approximation \cite{QSi01, Kotl06}. Compared to the Gross-Neveu-Yukawa-type model for the weak-coupling quantum criticality at high temperatures, two essential modifications have been made; the velocity of Z${}_{2}$ flux fluctuations is strongly renormalized to vanish and the dynamics of Ising spin excitations is governed by the inverse of the locally critical spin susceptibility instead of their relativistic dispersion.

Based on the empirical form of the spin susceptibility, Eq. (\ref{eq:chi_LQCP}), we write an effective action for the low-energy region as $S = S_{\mathrm K} + S_{\mathrm{AF}} + S_{\mathrm{K-AF}}$. This action is the same as Eq. (\ref{eq:eff_action}) (in Matsubara frequency space) except for $S_{\mathrm{AF}}$ which is modified as follows,
\begin{align}
&S_{\mathrm{AF}} = \frac{1}{2} \int_0^\beta\! d\tau d\tau' \int\!d^2x \ \phi(\tau,x) \chi_{\mathrm{LQCP}}^{-1}(\tau-\tau') \phi(\tau',x), \nn
&\chi_{\mathrm{LQCP}}^{-1}(i\omega, T, H) = \big[(aT + |\omega|)^\alpha + a^\alpha T^{*\alpha} \big]/A.
\end{align}
Note that $\chi_{\mathrm{LQCP}}^{-1}$ is inverse of Eq. (\ref{eq:chi_LQCP}). This expression looks quite similar to that for the description of high energy delocalized quantum criticality. However, there exist essential difference that both the spin and Z$_2$ flux dynamics are local. We perform the DMFT analysis in the non-crossing approximation \cite{Parc98,Mull84}, which confirms that this renormalization ansatz is self-consistent.

The one-loop self-energies for $\Psi$ and $X$ fermions are
\begin{align}
\Sigma_\Psi(i\omega,\bm k) = -\frac{g^2}{2}  \int_{i\Omega, \bm q} \chi_{\mathrm{LQCP}}(i\Omega) \mathcal G_X(i\omega - i\Omega, \bm k - \bm q), \nn
\Sigma_X(i\omega, \bm k) = -\frac{g^2}{2}  \int_{i\Omega, \bm q} \chi_{\mathrm{LQCP}}(i\Omega) \mathcal G_\Psi(i\omega - i\Omega, \bm k - \bm q),
\end{align}
where $\int_{i\Omega, \bm q} \equiv \frac{1}{\beta} \sum_{i\Omega} \int\!\frac{d^2q}{(2\pi)^2}$ and $\mathcal G_{\Psi, X}$ are renormalized Green's functions given by the Dyson equations
\begin{align}
\mathcal G_\Psi^{-1} =& i\omega \gamma^\tau + i\bm k \cdot \bm \gamma  - \Sigma_\Psi, \nn
\mathcal G_X^{-1} =& i\omega \gamma^\tau + i \bm M \cdot \bm \gamma - \Sigma_X.
\end{align}
Here, $\gamma^a$, $a = \tau, 1, 2$ are Euclidean form of Dirac gamma matrices in two-dimension. The self-consistent condition is formed from the renormalized spin susceptibility given by
\begin{align}
\chi_{\mathrm{LQCP}}(i\Omega) = \int_{i\omega, k} \Tr \big[ \mathcal G_X(i\omega, \bm k) \mathcal G_\Psi(i\omega + i\Omega, \bm k + \bm q) \big],
\end{align}
Resorting to self-consistent analyses in the non-crossing approximation, we found the susceptibility constraint for the power-law behavior with the critical exponent $\alpha $ as ${\chi }^{-1}(\omega ,T)\propto {\left(aT-i\omega \right)}^{\alpha }$, although a reliable $\alpha $-value is hard to be determined theoretically within the present mean-field analysis for the strong-coupling limit. Introduction of higher order perturbative corrections would yield the exponent $\alpha $ to be positive due to unitarity, in consistent with the phenomenological value $\approx 0.30$ determined from the INS results. These DMFT analyses at least confirm existence of the heavy-fermion-like local quantum criticality at low temperatures due to emergent local dynamics of the Z${}_{2}$ flux and Ising AFM fluctuations.

\section{Magnetic specific heat: Weak to Strong coupling crossover}
\label{sec:mag_sph}

\begin{figure}[t]
\includegraphics[width=8.5cm]{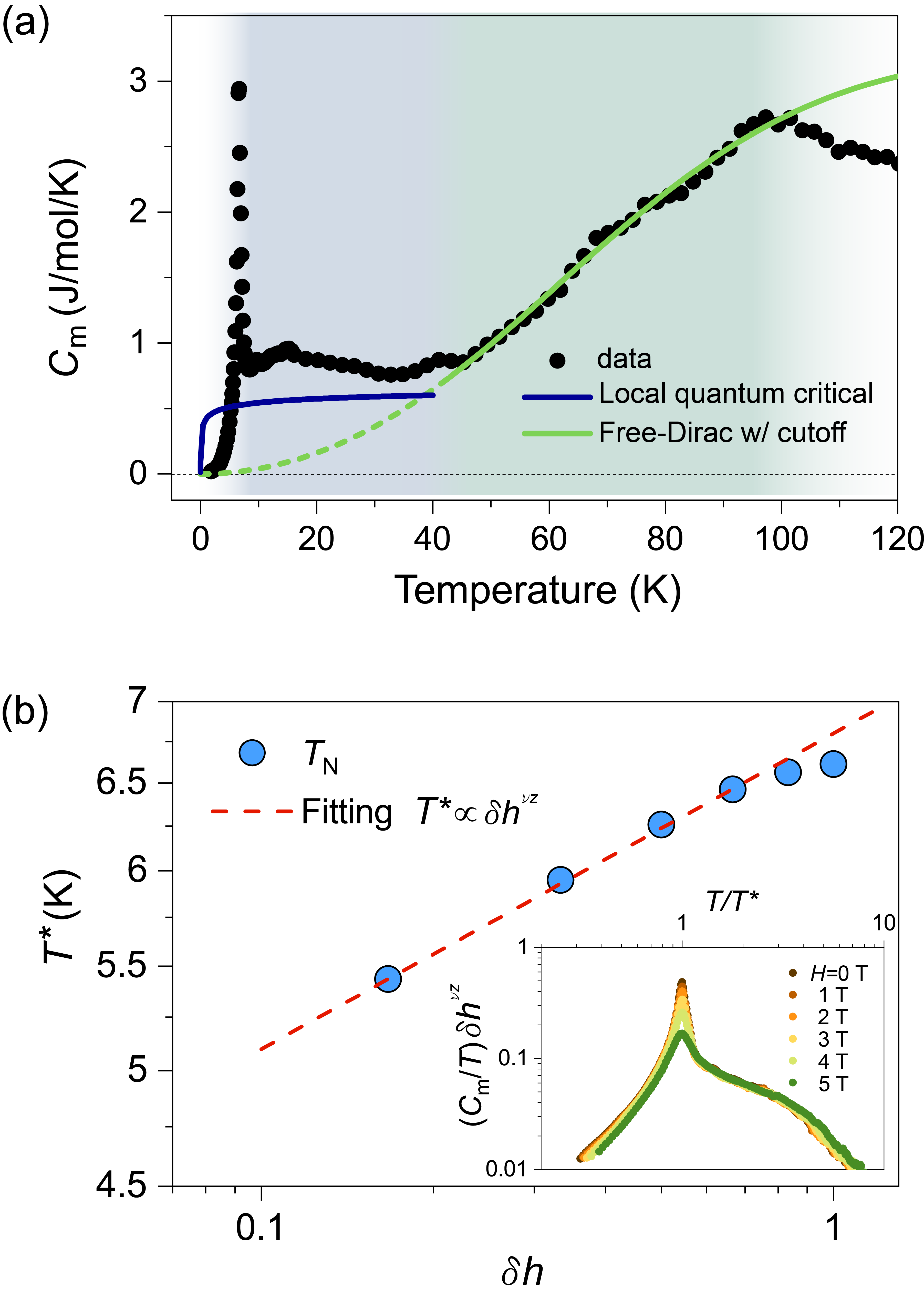}
\caption{Scaling behavior of the magnetic specific heat of \rucl{}.
(a) Magnetic specific heat (magnetic contribution) $C_{\rm m}$ obtained at the zero magnetic field and theoretical model calculations.
(b) Log-log plot of the N\'{e}el temperature $T_{\mathrm{N}}$ under magnetic fields $H$ as a function of $\delta h = 1-H/H_{\rm c}$ ($H \parallel a$) with the critical field $H_{\rm c} \approx 6\ \mathrm{T}$. The blue-filled circles denote $T_{\mathrm{N}}(H)$ extracted from the AFM peak position in $C_{\rm m}(H, T)$. The red-dashed line is given by scaling function for $T_{\mathrm{N}}\propto {\delta h}^{\nu z}$, where $\nu$ is the critical exponent of the correlation length and $z$ is the dynamical critical exponent. A universal scaling behavior of $C_{\rm m}$ is presented in the inset, where the scaling function data $\left(\frac{C_{\rm m}}{T}\right){\delta h}^{\nu z}$ for different $H$-fields merge into a single universal curve above $T_{\mathrm{N}}(H)$ with $\nu z \approx 0.125$ in the log-log plot as a function of $T/T_{\mathrm{N}}(H)$.}
\label{fig:scaling_sph}
\end{figure}

The crossover behavior of the quantum criticality is also observable in the magnetic specific heat $C_{\rm m}$ of \rucl{} \cite{SHDo17}. As shown in Fig. \ref{fig:scaling_sph}(a), $C_{\rm m}$ exhibits a low-temperature plateau up to $\sim 50\ \mathrm{K}$ above $T_{\mathrm{N}}$ and then follows a $T$-linear like behavior up to $T_{\mathrm{H}} \sim 100\ \mathrm{K}$. The latter, which was attributed to the Dirac-like itinerant MFs \cite{Nasu15}, indeed agrees well with the contribution calculated in a system with free Dirac fermions, in consistent with the weak-coupling quantum criticality. Remarkably, the plateau feature, which has been puzzling, turns out to be understood with the low energy local quantum criticality. Using the dynamic spin susceptibility, we construct a standard form of the free energy $F\left(T,H\right)$ = $-T\sum_{i\omega} \log \chi (i\omega, T, H)$ and calculate $C_{\rm m}\left(T, H = 0 \right)$ = $-T \partial^2F/\partial T^2$ with the fitting parameters $\alpha \approx 0.30$ and $a \approx 0.95$ obtained from the low energy scale INS results (see Supplemental Material B1 and Fig. S7 for details). As can be seen in the figure, the calculated $C_{\rm m}$ reasonably well reproduces the plateau feature above $T_{\mathrm{N}}$. This result implies that most of the entropy are given by the critically fluctuating local moments in this regime.

Besides the plateau feature, the strong-coupling local quantum criticality also predicts a scaling behavior of the specific heat $C_{\rm m}$ (also see Fig. \ref{fig:bulk_property} in Appendix \ref{sec_app:sph} and Supplemental Material B). Now the energy scale separated from QCP corresponds to $T_{\mathrm{N}}(H)$ = $T_{\mathrm{N}}(0) {\left(1-H/H_{\rm c}\right)}^{\nu z}$, inverse of the quantum coherence time at an $H$-field. The critical exponent $\nu z$ involves the critical exponent $\nu$ of the correlation length and the dynamic critical exponent $z$. In the local quantum criticality, $\nu \to 0$ (local) and $z \to \infty$ to yield a finite $\nu z$. To determine the exponent $\nu z$, we measured the specific heat $C_{\rm m}$ at various magnetic fields below the critical field $H_{\rm c} \approx 6\ \mathrm{T}$, where $T_{\mathrm{N}}(H_{\mathrm{c}})$ is supposed to become zero. Figure \ref{fig:scaling_sph}(b) shows $T_{\mathrm{N}}$ vs. $\delta h$ = $1-H/H_{\mathrm{c}}$ and the best fit is obtained with $\nu z \approx 0.125$. Using this value, we examine the scaling behavior of $\left(C_{\rm m}/T\right){\delta h}^{\nu z}$ as a function of $T/T_{\mathrm{N}}(H)$. As can be seen in the inset, all $\left(C_{\rm m}/T\right){\delta h}^{\nu z}$ at different $H$-fields merge into a universal scaling curve, indicating that the effective spatial dimension, in which critical fluctuations dominate the entropy-carrying, is extremely local. This result confirms that the low temperature specific heat above $T_{\mathrm{N}}$ is governed by the strong-coupling local quantum criticality.

\section{Summary and discussions}
\label{sec:summary}

\subsection{Summary}

In this study, we show that \rucl{} crossover from the high-energy weak-coupling critical region to the low-energy strong-coupling critical region above $T_{\rm N}$ as described in Fig. \ref{fig:phase_diagram}. We identified a Wilson-Fisher-Yukawa FP which governs the universal physics in the high-energy region. It is essentially the same physics with that of the pure Kitaev model. Although the zigzag Ising AFM fluctuations are introduced in the present effective field-theory description, they just contribute short-ranged effective interactions to both matter fluctuations, itinerant MFs and localized Z$_2$-flux excitations, at the high temperature quantum critical regime. In this respect, it is not surprising to have the reasonable agreement between the experiment and the pure Kitaev theory, and the essential ingredients in our field theory description are almost the same as those of the simulation from the pure Kitaev model \cite{Nasu15, Yosh16, Bane17,SHDo17}. Here, we want to emphasize that the present study on the Kitaev-AFM model determines the explicit formula for the scaling function, Eq. (\ref{eq:scaling_highE}), which cannot be obtained from previous numerical studies.

The more interesting discovery is in the low-energy, strong-coupling region. We demonstrated experimentally and theoretically that the spin dynamics follows the heavy-fermion-type strong-coupling physics at low energies. We could show this emergent strong-coupling local quantum criticality at low energies appears from the weakly-coupled rather conventional quantum criticality at high energies. This weak-coupling (Wilson-Fisher type) to strong-coupling (locally critical heavy-fermion type) quantum criticality crossover revealed in \rucl{} has not been expected before. The local quantum criticality is cross-checked both theoretically and experimentally; self-consistent analysis based on the DMFT description and the magnetic specific heat measurement, respectively. The scaling behavior at low temperatures above $T_{\rm N}$ indicates that the critical fluctuations dominating the entropy-carrying is extremely local. Unfortunately, we could not explicitly derive the crossover regime since it requires a non-perturbative theoretical approach, which is quite complicated and too difficult. Instead, we verified existence of the heavy-fermion like strong-coupling FP in a self-consistent way based on the DMFT. Again, we point out that this strong coupling phenomena has never been either observed or discussed in the research of \rucl{}.

\subsection{Comparison with heavy-fermion quantum criticality}

Note that the existence itself of two FPs is not surprising. For an example, suppose a scalar field theory with an effective $\phi^4$-type interaction, regarded to be an effective field theory for a transverse-field Ising lattice model \cite{KardarBook}. This field theory is well known to show its RG flow from a non-interacting Gaussian FP at the high-energy UV regime to an interacting Wilson-Fisher one at the low-energy IR regime \cite{PeskinBook}. In quantum chromodynamics, there is an RG flow from an (“almost” non-interacting) asymptotically free theory to a strong-coupling confinement phase \cite{PeskinBook}. Here, we have an RG flow between the weak-coupling Wilson-Fisher-type “conventional” UV FP and the strong-coupling heavy-fermion-type “unconventional” or local IR FP. A remarkable point is that we reveal the nature of the IR FP in this Kitaev-type material \rucl{}. In particular, this IR FP is strongly correlated to be locally quantum critical. Emergence of this heavy-fermion-type strong-coupling FP in this material is completely unexpected. 

One may criticize that the heavy-fermion system shows a similar weak-coupling to strong-coupling quantum criticality crossover near the heavy-fermion magnetic QCP. Indeed, some crossover behaviors have been observed in thermodynamics and transport measurements \cite{Gege05}. However, these crossover behaviors were not clearly understood both experimentally and theoretically. For examples, there is a crossover behavior in the specific heat of YbRh$_2$Si$_2$ near a magnetic-field tuned QCP \cite{Gege05}. One may claim that the high-temperature region would be governed by the Hertz-Moriya-Millis theory, a standard weak-coupling theory for heavy-fermion quantum criticality \cite{Zhu03}. Meanwhile, there is no consensus for the low-temperature anomalous behavior, not understood as far as we know. In addition, there is a classical paper on INS measurements for CeCu$_{\mathrm{(6-x)}}$Au$_{\mathrm x}$~\cite{QSi01}, in which the low-energy spin dynamics was described in a momentum-independent local form with $\omega/T$ scaling. The origin of this functional form was proposed based on a DMFT framework, but the high-energy spin dynamics was not clarified in the study and it has not been understood yet how the spin dynamics evolves from UV to IR.

Analogous to the crossover between the weak-coupling Hertz-Moriya-Millis quantum criticality and the strong-coupling local quantum criticality in heavy-fermion systems, we verified a similar quantum criticality crossover behavior in \rucl{} not only theoretically but also experimentally. Our study verifies the quantum criticality crossover behavior in \rucl{}. A similar crossover was suggested in the heavy-fermion system, but its mechanism has not been understood in our opinion.

\subsection{Perspectives}

If one could simulate the Kitaev-Heisenberg lattice model \cite{Jack09} and calculate the spin-spin correlation function in a brute force way, we believe that the low-energy strong coupling quantum critical physics can be verified theoretically. It remains as an interesting future research. In this respect, an interesting message given by the present study is that the Kitaev-Heisenberg lattice model (with a gamma term) would show the weak-coupling to strong-coupling quantum criticality crossover behavior from UV to IR. To confirm the locally quantum critical scaling function at low energies more transparently, we have to show momentum independence of the low-energy spin spectrum. In other words, we have to investigate the scaling plot at other transfer momentum points in our neutron scattering measurements. In addition, we can calculate both longitudinal and transverse thermal conductivities based on this locally quantum critical scaling function for the spin spectrum. Resorting to the DMFT framework with this local spin spectrum, we calculate the self-energy of the itinerant fractionalized fermion excitations, which gives the temperature dependence of their scattering rate. Here, the main point is that localization of Z$_2$ gauge fluxes causes that of Ising antiferromagnetic fluctuations, both of which are coupled to the delocalized fractional excitations. As a result, we suspect that both thermal transport coefficients would show effectively a metal (UV) to insulator (IR) crossover behavior due to the localization physics. We expect that this physics may be reflected in a $H/T$ scaling function for the thermal conductivities.

We believe that our discovery of the crossover behavior from deconfined weak-coupling `non-local' quantum criticality to deconfined strong-coupling `local' quantum criticality in \rucl{} opens a new research field of critical quantum spin liquids, which result from the interplay between the spontaneous symmetry breaking and the topological ordering. In particular, thermal transport properties in \rucl{} would reveal novel transport phenomena distinguishable from existing ones, giving rise to a new universal scaling law of the transport properties due to fade-out of well-defined fractionalized excitations in spite of the topological ordering. In this aspect, \rucl{} provides an ideal platform to explore a novel universality class, where new universal scaling laws govern the thermodynamic, spectroscopic, and transport properties.

\begin{acknowledgments}
This work is supported by the Max Planck POSTECH/Korea Research Initiative, Study by Nano Scale Optomaterials and Complex Phase Materials (2016K1A4A4A01922028), through the National Research Foundation (NRF) funded by MSIP of Korea. K.-S.K. and J.-H.H. acknowledge support from the NRF grant (NRF-2021R1A2C1006453 and NRF-2021R1A4A3029839). S.J. acknowledges support from NRF grants (NRF-2017M2A2A6A01071297 and NRF-2017R1D1A1B03034432). S.-H.D. and J.-Y.K. acknowledge partial support from a NRF grant (NRF-2017K1A3A7A09016303). We thank J. Ross Stewart and D. J. Voneshen for technical supports in the INS experiments and E.-G. Moon for fruitful theoretical discussions.
\end{acknowledgments}

\appendix


\section{EXPERIMENTAL DETAILS} \label{experimental_detail}
\begin{figure}[t]
\includegraphics[width=8.5cm]{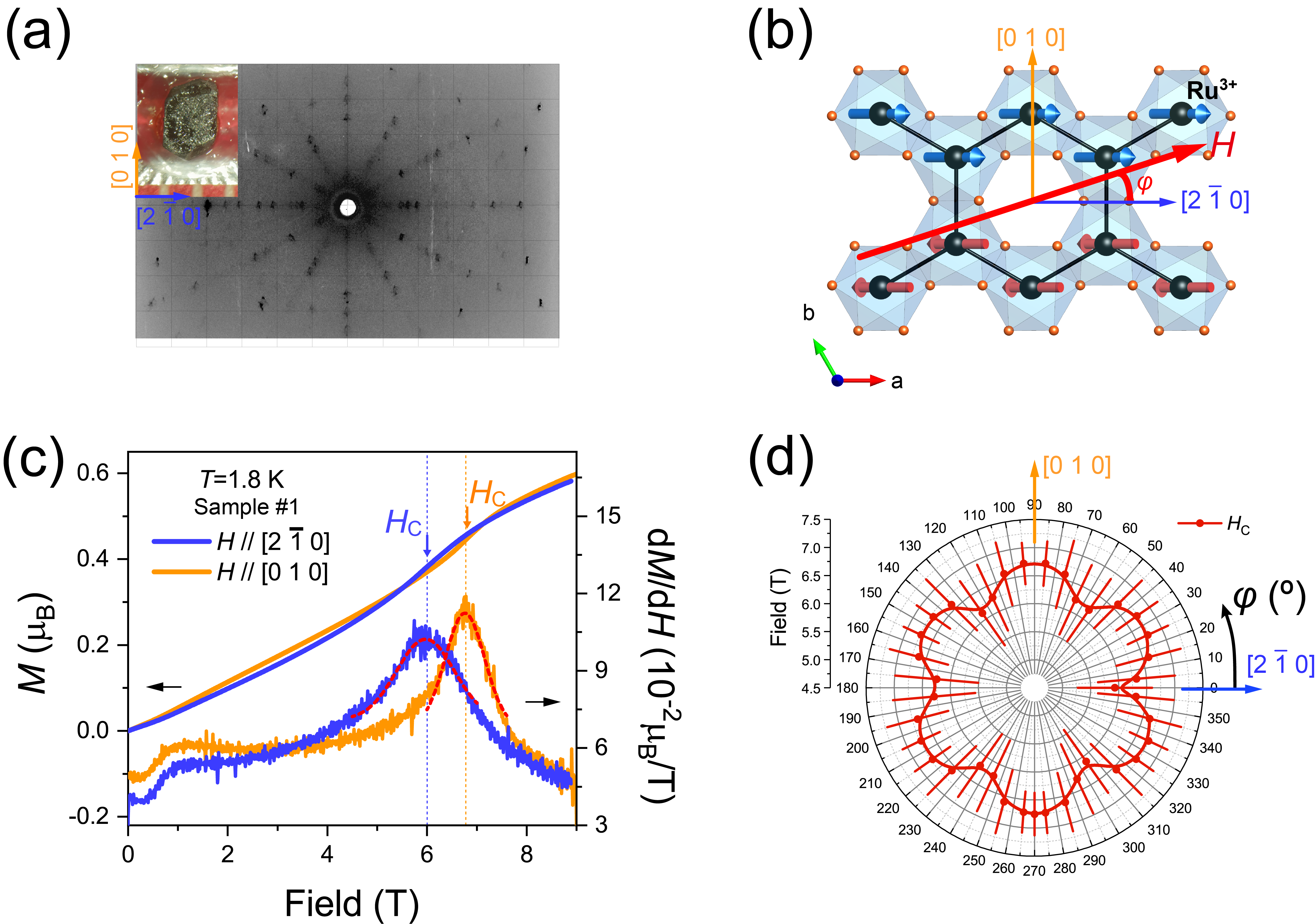}
\caption{Laue pattern and field-dependent magnetizations of \rucl{}.
(a) Laue image view perpendicular to the $ab-$plane. The Laue pattern reflects the near-hexagon structure at room temperature. Inset displays a picture of a single crystal (sample \#1) with crystal axes, used for the measurement.
(b) Local spin structure of the zigzag AFM order in \rucl{}. Two perpendicular vectors [2,  $\bar{1}$, 0] and [0, 1, 0] in hexagonal reciprocal space ($R3$ or $P3$) are presented on the real-space honeycomb lattice.
(c) Magnetizations and first derivatives with respect to $H$ along $H$ $\parallel$ [2,  $\bar{1}$, 0] and $H$ $\parallel$ [0, 1, 0] at $T$ = 1.8 K. Red dashed lines present Gaussian fits for the first derivative curves, which exhibit single anomalies at $H_{\mathrm{C}}$ = 6.0~T and $H_\text{C}$ = 6.8~T for $H$ $\parallel$ [2, $\bar{1}$, 0] and $H$ $\parallel$ [0, 1, 0] as marked by blue and orange arrows, respectively.
(d) Polar plot for the angular dependence of $H_\text{C}$. The azimuthal angle $\varphi$ of the $H$-direction is referred to [2, $\bar{1}$, 0] in the polar plot. The dot and error bar indicate $H_\text{C}$ and the width of the Gaussian fit for $dM/dH$ as shown in (c), respectively.}
\label{fig:bulk_property}
\end{figure}

Single crystalline \rucl{} samples were prepared using the vacuum sublimation method as described in Ref. \cite{SHDo17}. The crystal orientation was determined by using the X-ray Laue (Fig. \ref{fig:bulk_property} (a)). The magnetization and specific heat of \rucl{} were measured by using the conventional vibrating sample magnetometry (VSM) and calorimeter equipped at a commercial Quantum Design Physical Property Measurement System (PPMS-Dynacool), respectively. The field dependent magnetization and specific heat results are displayed in Fig. \ref{fig:bulk_property} and Fig. \ref{fig:spH_mag_suppl}.


Zero-field INS measurements were performed by using MERLIN and LET time-of-flight spectrometers at the ISIS spallation neutron source in Rutherford Appleton Laboratory, UK. 165 pieces of \rucl{} single crystals with the total mass $\sim$ 5.5 g were co-aligned on an aluminum plate with the (0 \textit{K L}) scattering plane. The samples were placed under a liquid helium flow-type cryostat with the temperature control ranging from 2 K to 290 K. In the MELRIN experiments, we used a chopper frequency of 300 Hz, which provides $E_\text{i}$ = 12 meV, 22 meV, and 50 meV of incident neutron energies with FWHM (full width at the half maximum) energy resolutions of 0.34 meV, 0.75 meV, and 2.23 meV at elastic scattering, respectively. The measurements were performed for the sample rotation from -52$^\circ$ to 52$^\circ$ with 4$^\circ$ step referring to 0$^\circ$ at $k_i \parallel c^*$. The LET experiments were performed at $E_\text{i}$ = 10 meV with the energy resolution of 0.36 meV (FWHM) for the elastic scattering, and the sample rotation from -30$^\circ$ to 30$^\circ$ with 5$^\circ$ step relative to $k_i \parallel c^*$. The background signals were separately obtained by using an identical aluminum sample holder both in the MERLIN and LET experiments for the background correction. INS data were normalized, and converted to the unit for the neutron scattering function by using the incoherent neutron scattering intensity of a standard reference vanadium sample. 

The data presented in Figs. \ref{fig:spectra_high} and \ref{fig:spectra_low} were obtained by integration over [0, 0, \textit{L}] = [-2.5, 2.5], [0, \textit{K}, 0] = [-0.17, 0.17], and [\textit{H}, -0.5\textit{H}, 0] = [-0.17, 0.17]. The magnetic form factor contribution from the \textit{L}-component in the integrated data was corrected by dividing with the Ru${}^{3+}$ magnetic form factor at each data point \cite{SHDo17}, and the scaled data were compared with the theoretical model calculations for $S(\mathrm{\Gamma}, \omega)$. All the data were analyzed by using HORACE software distributed by ISIS \cite{Ewin16}.

\section{THEORETICAL ANALYSIS FOR WEAK- AND STRONG-COUPLING QUANTUM CRITICALITY}
\label{theoretical analysis}

\subsection{Weak-coupling quantum criticality}

We constructed an effective coarse-grained lattice model Hamiltonian consisting of the Kitaev QSL Hamiltonian $H_{\mathrm{K}}$, the zigzag AFM Ising spin Hamiltonian $H_{\mathrm{AF}}$, and an effective Zeeman-type interaction term $V_{\mathrm{K-AF}}$ accounting for coupling between the Kitaev and the zigzag ordered Ising spins. Performing the Jordan-Wigner transformation, we obtained an effective lattice Hamiltonian, which presents \textit{p}-wave superconducting-type paired itinerant fermions interacting with localized fermions of the Z${}_{2}$ flux fluctuations. Taking the continuum limit, we found an effective field theory in a form of the Gross-Neveu-Yukawa type theory. Performing the Wilsonian RG analysis in the one loop level, we revealed existence of a Wilson-Fisher-Yukawa-type novel FP, which corresponds to the high-energy scale weak-coupling FP, as described in the RG flow diagram presented in Fig. \ref{fig:phase_diagram}(b) and Fig. \ref{fig:RG_flow}. Solving the RG equation for the dynamic spin susceptibility near this FP, we derive the theoretical universal scaling function of $\chi'' (\Gamma ,\omega )T$ with a single free parameter of an overall scaling coefficient $\chi_0$, and fit the experimental results scaled in $\hbar\omega / k_B T$, which are determined from the INS spectra in a high energy range of 4 meV $\le \hbar\omega \le$ 15 meV as shown in Fig. \ref{fig:spectra_high} (log-log scale). The linear scale comparisons between the experimental and theoretical $\chi''(\Gamma,\omega)$s for different temperatures are also presented in Fig.~\ref{fig:neutron_data_appnd}. The theoretical details are presented in Supplemental Material A.

\subsection{Strong-coupling quantum criticality}

This universal scaling behavior of the dynamic spin susceptibility changes drastically below around 4meV, and the experimental $\chi''(\mathrm{\Gamma}, \omega ) T^{0.3}$ values in a low energy range of 1 meV $\le \hbar\omega \le$ 5 meV were remarkably well fitted by using the following local form of the dynamic spin susceptibility $\chi \left(\omega ,T,H\right) = \frac{A}{{\left(aT-i\omega \right)}^{\alpha }+a^{\alpha }T^{*\alpha }}$ with $\alpha \approx 0.30$, $A \approx 0.144$, and $a \approx 0.95$ as shown in Fig. \ref{fig:spectra_high}. To confirm the existence of this heavy-fermion-like local quantum criticality in the low energy scale, we perform a DMFT analysis in the non-crossing approximation. In the analysis, we found that the velocity of Z${}_{2}$ flux fluctuations is renormalized strongly enough to approach to a zero-value limit and the dynamics of Ising spin excitations is governed by the inverse of this locally critical spin susceptibility instead of their relativistic dispersion. As a result, we confirm the existence of the local quantum criticality at low energies due to emergent local dynamics of both Z${}_{2}$ flux and Ising-type AFM fluctuations, and explain why the power-law critical exponent $\alpha \approx 0.3$ in $T^{-\alpha }$ of the dynamic spin susceptibility is much smaller than that of the independent magnetic local moments, i.e. $T^{-1}$ Curie law behavior. The theoretical details are presented in Supplemental Material B.

\section{COMPARISON BETWEEN INELASTIC NEUTRON SCATTERING AND THEORETICAL MODELS}

\begin{figure}[t]
\includegraphics[width=8.5cm]{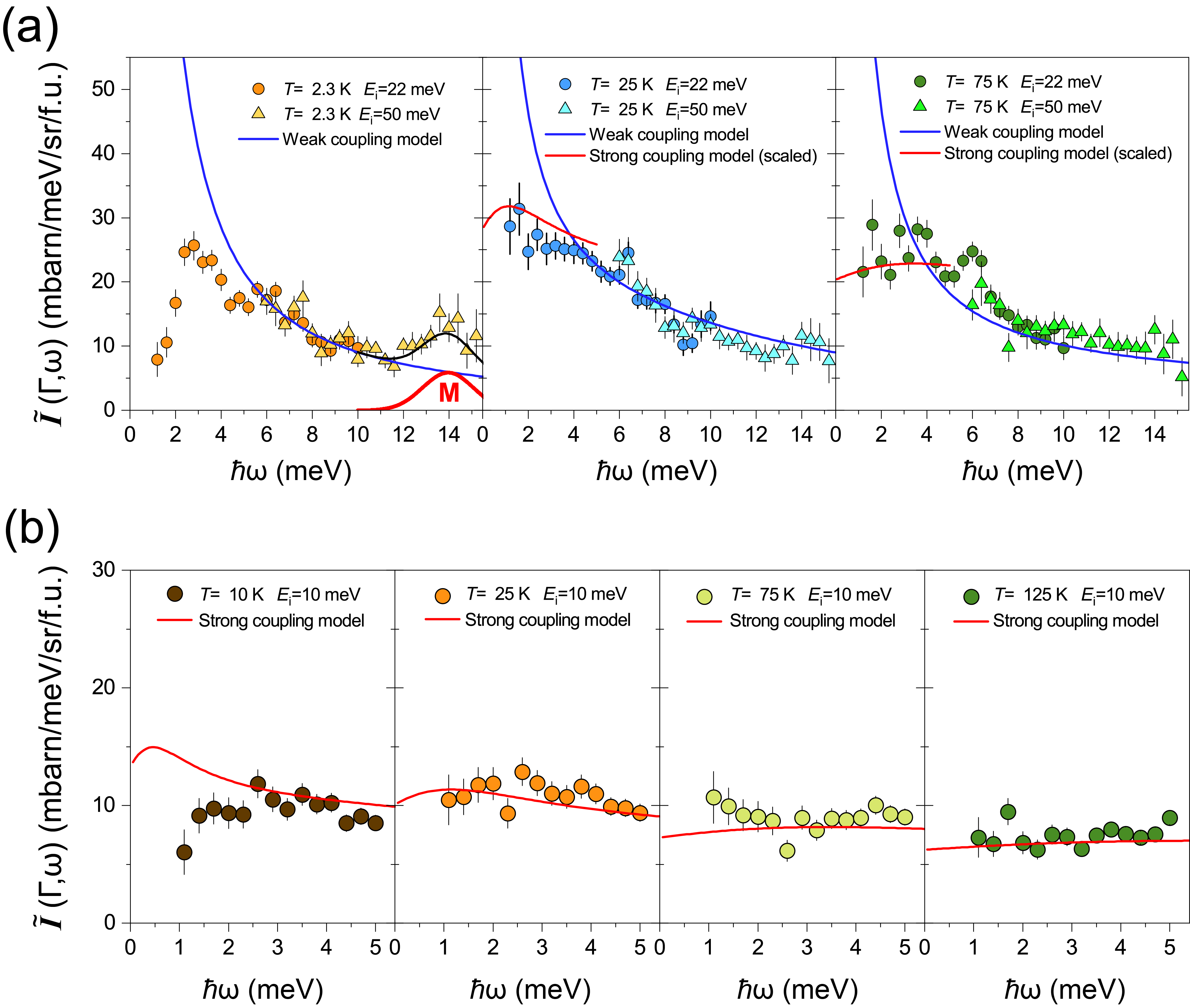}
\caption{Comparison of the inelastic neutron scattering with the theoretical models. (a) Neutron scattering function $S(\Gamma,\omega)$s measured from MERLIN experiment by using $E_\text{i}$=22~meV and 50~meV, with integration over [0,0,$L$]=[-2.5, 2.5], [0,$K$,0]= [-0.17,0.17], and [$H$,-0.5$H$,0]=[-0.17,0.17].
The theoretical scaling functions are extracted from the weak coupling model (blue line) for a high energy scale of 4 $\mathrm{\le}$ $\hbar \omega$ $\mathrm{\le}$15 meV and the strong coupling (red line) for a low energy scale of 0 $\mathrm{\le}$ $\hbar \omega$ $\mathrm{\le}$5~meV. Below $T_\text{N}$ (2.3~K), only the high energy weak coupling model is presented in the figure since the low energy excitations are dominated by the AFM spin wave excitations which are not taken into account in the low energy strong coupling model. The AFM multi-magnon contributions (red solid line), marked with `M', are presented by a Gaussian function centered at 14~meV. (b) $S(\Gamma,\omega )$ from LET experiment using $E_{i}$=10~meV, integrated over [0,0,$L$]=[-2.5,2.5], [0,$K$,0]= [-0.17,0.17], and [$H$,-0.5$H$,0]=[-0.17,0.17].
The obtained data (filled circle) are compared with the strong coupling model (solid red line).}
\label{fig:neutron_data_appnd}
\end{figure}

Figure \ref{fig:neutron_data_appnd}(a) shows neutron scattering functions $S(\bm Q, \omega)$ for $\bm Q = \Gamma$ (0, 0, 0) at temperatures $T$ = 2.3 K, 25 K, and 75 K. The data exhibit the continuum excitations emerging on cooling across $T_{\mathrm{H}}$ ($\sim$ 100 K), and the continuum still remains even below $T_{\mathrm{N}}$ $\sim$ 6.5 K. The experimental $S(\Gamma, \omega)$s are compared with the theoretical spectra in the linear-linear scale. The theoretical spectra, which are separately extracted from two different models, the weak coupling model (Eq.~(\ref{eq:scaling_highE}) in the main text) and strong coupling model (Eq.~(\ref{eq:scaling_lowE}) in the main text), are presented by blue and red lines in the high and low energy ranges of 4 meV $\le \hbar\omega \le$ 15 meV and 1 meV $\le \hbar\omega \le$ 5 meV, respectively. The models well reproduce the respective overall line shapes of $S(\Gamma, \omega)$s in both energy ranges except additional contributions, which are not included in the model spectra. In the comparison for $S(\Gamma,\omega)$ at $T$ = 2.3 K (blow $T_{\mathrm{N}}$), we only present the weak coupling model spectrum in the high energy range since $S(\Gamma,\omega)$ in the low energy range is strongly disturbed by the apparent AFM spin wave excitations, which are not taken into account in the spectrum of the low energy strong coupling model. A broad peak, which involves the AFM multi-magnon modes, also appears around 14 meV. The spectral weight of the multi-magnon contribution was simply formulated with a phenomenological Gaussian function (red solid line) as shown in the left first figure. The phonon contribution, which increases upon heating, also appears as an additional sharp peak at $\hslash \omega \simeq 6\ \mathrm{meV}$ \cite{SHDo17, HLi20}, and causes a deviation from the weak coupling model.

Figure \ref{fig:neutron_data_appnd}(b) presents a detailed comparison of the $S(\Gamma, \omega)$s in the low energy region, 1 meV $\le \hbar\omega \le$ 5 meV, with the strong coupling model. The calculated spectra from the strong coupling model reproduce the $S(\Gamma, \omega)$s at different temperatures above $T_{\mathrm{N}}$ up to 125 K.

\begin{figure}[t!]
\includegraphics[width=8.5cm]{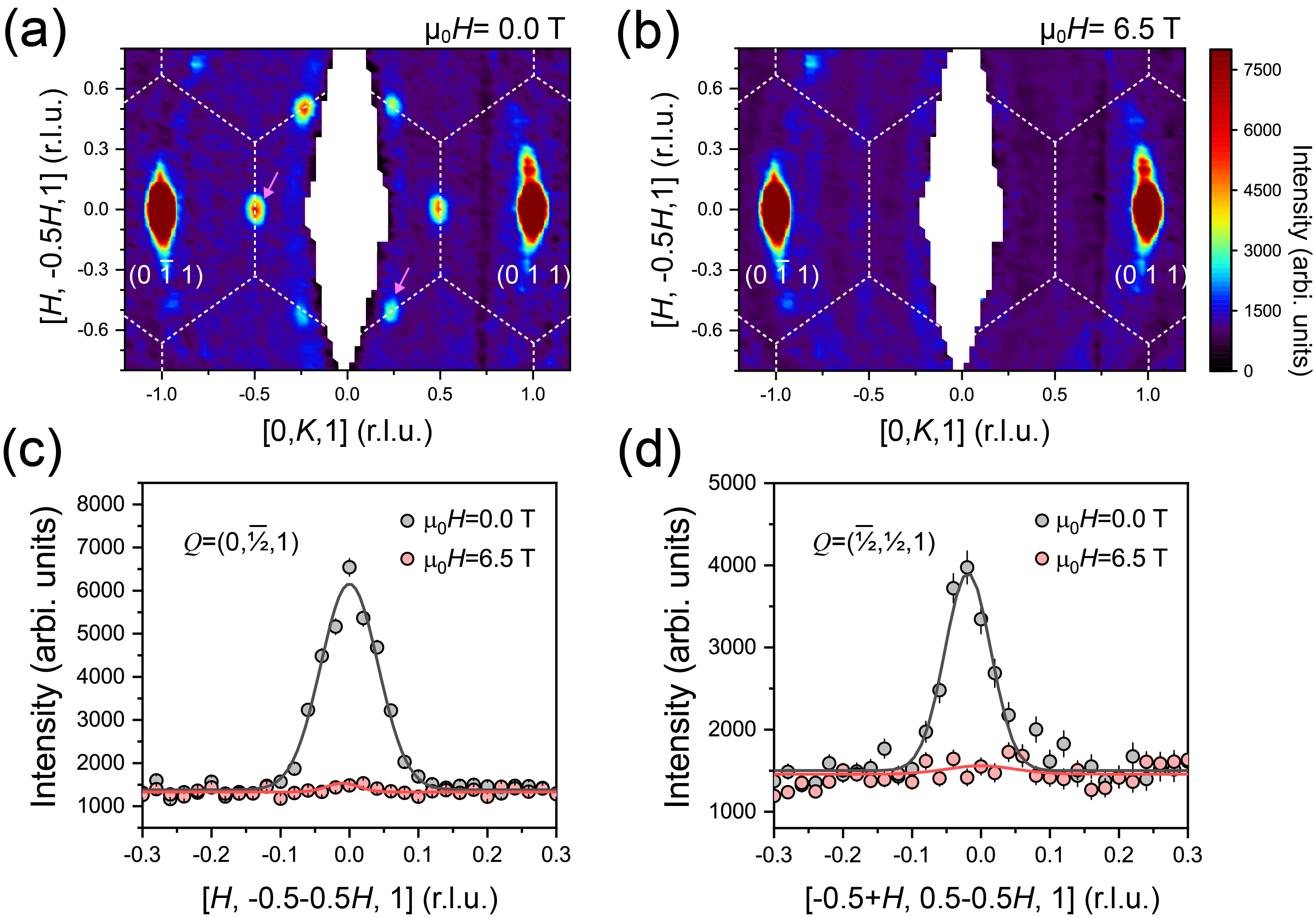}
\caption{Elastic neutron scattering results of \rucl{} single crystals at $T$ = 1.8 K. Elastic neutron scattering intensity maps at \textit{L} = 1 on the \textit{HK-}plane with $H$ = 0 T (a) and $H$ = 6.5 T (b) along $[2, \bar 1, 0]$. The scattering data were obtained with $E_\text{i}$ = 22 meV by using the LET spectrometer. Two representative magnetic Bragg reflection peaks for $H$ = 0 T and $H$ = 6.5 T at $\bm Q = (0, \bar{1\over2}, 1)$ (c) and $(\bar{1\over2}, {1\over2}, 1)$ (d). The peak positions are indicated with pink arrows in the $H$ = 0 T map. The peak intensities are mostly suppressed at $H$ = 6.5 T, meaning that the long-range zigzag AFM order parameter nearly vanishes.}
\label{fig:neutron_data_mag_suppl}
\end{figure}

\section{MAGNETIZATION AND NEUTRON DIFFRACTION IN MAGNETIC FIELDS}

\begin{figure*}[t]
\includegraphics[width=17cm]{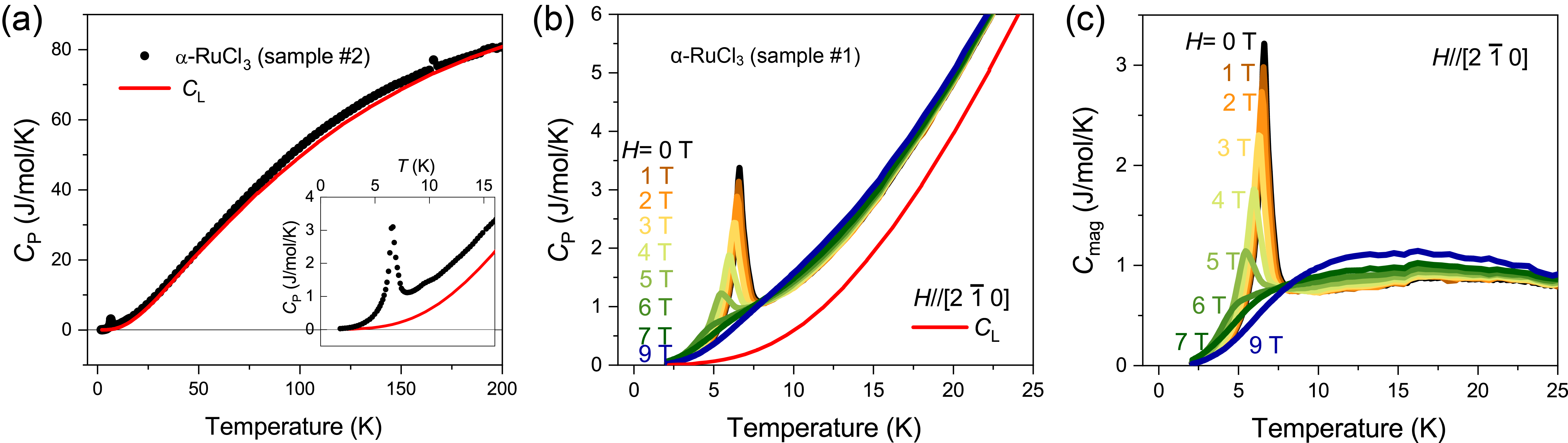}
\caption{Specific heat measurements in magnetic fields.
(a) Specific heat of \rucl{} and the lattice contribution $C_L$ (red solid line) estimated from ScCl${}_{3}$. Inset shows a magnification of specific heats at a low temperature region. A black arrow indicates the AFM transition peak at $T_{\mathrm{N}} \simeq$ 6.5 K.
(b) Specific heat of \rucl{} for different $H$-fields along $[2, \bar 1, 0]$ ranging from 0 to 9 T. \textit{C}${}_{L}$ is presented by a red solid line again.
(c) Magnetic specific heat of \rucl{} for the different $H$-fields.}
\label{fig:spH_mag_suppl}
\end{figure*}

The crystal axes of \rucl{} single crystal (sample \#1) were determined by using the Laue diffraction (Fig. \ref{fig:bulk_property}(a)). The magnetization was measured under applied magnetic fields in the \textit{ab-}plane (see Fig. \ref{fig:bulk_property}(b)). Figure \ref{fig:bulk_property}(c) shows the magnetizations and their first derivatives under the in-plane fields of $H \parallel [\bar 2 \ 1 \ 0]$ and $H \parallel [0 \ 1 \ 0]$ in the \textit{HK-}reciprocal space at 1.8 K. The first derivatives exhibit magnetic anomaly peaks, which represent the field-induced transition from the zigzag AFM to a partially polarized spin state. The anomalies were fitted with Gaussian functions, which yield the critical fields $H_{\mathrm{C}} \simeq$ 6 T and 6.8 T along $H \parallel [\bar 2 \ 1 \ 0]$ and $H \parallel [0 \ 1 \ 0]$, respectively. To investigate detailed angular dependence of $H_{\mathrm{C}}$, magnetization was measured with a magnetic field rotating from 0$^\circ$ to 360$^\circ$ with 5$^\circ$ step in the \textit{ab-}plane. The angular critical fields presented in Fig. \ref{fig:bulk_property}(d) reveal the six-fold symmetry of $H_{\mathrm{C}}$ oscillating from 6 T to 6.8 T, reflecting the $C_3$ rotational symmetry of the honeycomb lattice in \rucl{}. $H_{\mathrm{C}}$ becomes the minimum (maximum) value at 6 T (6.8 T) for the $H$-direction line lying at the honeycomb edge (vertex).

Figure \ref{fig:neutron_data_mag_suppl} shows elastic neutron scattering results at $T$ = 1.8~K in the \textit{HK-}reciprocal space without and with $H \parallel [2 \ \bar 1 \ 0]$ (honeycomb edge) slightly above $H_{\mathrm{C}}$. At $H=0$, sharp magnetic Bragg peaks are clearly shown at $(1 \ 0 \ {1\over2})$ corresponding to the zigzag AFM order. The presence of magnetic domains with 120$^\circ$ and 240$^\circ$ in-plane rotations produces six-magnetic Bragg peaks at the M-points in the first Brillouin zone. At $H$ = 6.5~T ($> H_{\mathrm{C}} \approx$ 6 T), all the magnetic Bragg peaks nearly disappear, indicating absence of the long-range AFM order in consistent with the magnetization result in Fig. \ref{fig:bulk_property}(c).

\section{SPECIFIC HEAT OF \rucl{}}
\label{sec_app:sph}

To study the thermodynamic behavior of \rucl{}, we performed specific heat ($C_P$) measurements the single crystalline sample. Figure \ref{fig:spH_mag_suppl}(a) shows the temperature dependence from 1.8 K to 200 K at $H$ = 0 T and Fig. \ref{fig:spH_mag_suppl}(b) does the temperature dependence in a 1.8 K to 25 K window for different $H$-fields ($H \parallel [2 \ \bar 1 \ 0]$) ranging from 0 T to 9 T. The specific heat $C_P$ of \rucl{} consists of magnetic ($C_{mag}$) and lattice ($C_L$) contributions. $C_L$, which is presented by a red solid line in both figures, was estimated from $C_P$ of a non-magnetic iso-structural compound ScCl${}_{3}$ with Debye temperature scaling \cite{Bouv91}. Figure \ref{fig:spH_mag_suppl}(c) displays the field dependent magnetic specific heat $C_{mag}$ obtained by subtracting $C_L$ from $C_P$ ($C_{mag} = C_P - C_L$) of \rucl{}.

The zero-field $C_{mag}$ shows a sharp peak at $T_{\mathrm{N}} \simeq$ 6.5 K, and exhibits the plateau behavior above $T_{\mathrm{T}}$ as expected in the local quantum critical region described in the main text. With increasing magnetic field, the peak position and height gradually decrease, and the peak disappears across $H_{\mathrm{C}} \simeq$ 6 T. We note that the transition temperature $T_{\mathrm{N}}(H)$ is not well defined at 6 T near the critical field.

\providecommand{\noopsort}[1]{}\providecommand{\singleletter}[1]{#1}%
%


\end{document}